\documentclass[11pt]{article}
\usepackage[latin9]{inputenc}
\usepackage{geometry}
\usepackage{amsmath}
\usepackage{amssymb}
\usepackage{stmaryrd}
\usepackage{graphicx}
\usepackage{setspace}
\usepackage[authoryear]{natbib}
\usepackage[unicode=true,
 bookmarks=false,
 breaklinks=false,pdfborder={0 0 1},backref=section,colorlinks=false]
 {hyperref}
\hypersetup{
 colorlinks,citecolor=blue,pdftex}
\usepackage{todonotes}
\setuptodonotes{size=\small} 
\makeatletter

\usepackage{amsfonts}
\usepackage{amsthm}
\usepackage{lscape}
\usepackage{appendix}
\usepackage{dcolumn}
\usepackage{rotating}
\usepackage{comment}
\usepackage{color}
\usepackage{float} 

\usepackage{pgf}
\usepackage{tikz}\usepackage{caption}
\usepackage{subcaption}
\usepackage{tkz-tab}
\usepackage{tikz-3dplot}
\usetikzlibrary{calc}
\usetikzlibrary{arrows, automata}

\newcolumntype{d}[1]{D{.}{.}{#1}}
\newcolumntype{t}[1]{D{,}{,}{#1}}
\newcolumntype{i}[1]{D{.}{}{#1}}
\newtheorem{theorem}{Theorem}[section]

\newtheorem{corollary}{Corollary}[section]

\newtheorem{definition}{Definition}[section]
\newtheorem{example}{Example}
\newtheorem{lemma}{Lemma}[section]

\newtheorem{remark}{Remark}[section]

\theoremstyle{plain} 

\newtheorem{innercustomthm}{Condition}
\newenvironment{mycondi}[1]
  {\renewcommand\theinnercustomthm{#1}\innercustomthm}
  {\endinnercustomthm}

\newtheorem{innercustomas}{Assumption}
\newenvironment{myas}[1]
  {\renewcommand\theinnercustomas{#1}\innercustomas}
  {\endinnercustomas}

\usepackage{bbm}

\numberwithin{equation}{section}

\makeatother

\begin{document}
\title{On Quantile Treatment Effects, Rank Similarity,\\
and Variation of Instrumental Variables\thanks{For insightful discussions, the authors are grateful to Victor Chernozhukov and Xinyue Bei, and also to participants at the Econometric Society North American Winter Meeting and Asian Meeting (2023), the Barcelona School of Economics Summer Forum, the CeMMAP-SNU Conference, the 2024 CUHK Econometrics Workshop, the 2025 CES North America Annual Conference, Econometrics Mini-conference at UCLA, and seminars at UCL, the University of Oxford, Boston University, Boston College, the University of Iowa, Nanyang Technological University, and Singapore Management University. We also thank Xiaodan Fang for excellent research assistance. All errors are our own.}
}

\author{Sukjin Han\\
 School of Economics\\
 University of Bristol\\
\UrlFont{\sffamily}\href{mailto:sukjin.han\%5C\%40gmail.com}{sukjin.han@gmail.com}\and
Haiqing Xu\\
Department of Economics \\
 University of Texas at Austin \\
\UrlFont{\sffamily}\href{mailto:h.xu\%5C\%40austin.utexas.edu}{h.xu@austin.utexas.edu}}
\date{\today}

\maketitle
\vspace{-0.2cm}

{\hspace{100pt}\small Rewriting in progress: please excuse the debris!}

\begin{abstract}
This paper develops a nonparametric framework to identify and estimate distributional treatment effects under nonseparable endogeneity. We begin by revisiting the widely adopted \emph{rank similarity} (RS) assumption and characterizing it by the relationship it imposes between observed and counterfactual potential outcome distributions. The characterization highlights the restrictiveness of RS, motivating a weaker identifying condition. Under this alternative, we construct identifying bounds on the distributional treatment effects of interest through a linear semi-infinite programming (SILP) formulation. Our identification strategy also clarifies how richer exogenous instrument variation, such as multi-valued or multiple instruments, can further tighten these bounds. Finally, exploiting the SILP's saddle-point structure and Karush-Kuhn-Tucker (KKT) conditions, we establish large-sample properties for the empirical SILP: consistency and asymptotic distribution results for the estimated bounds and associated solutions.



\end{abstract}

\section{Introduction\label{sec:Introduction}}
This paper develops a nonparametric framework to identify and estimate distributional treatment effects under nonseparable endogeneity. We begin by revisiting the widely adopted \emph{rank similarity} (RS) assumption and characterizing it by the relationship it imposes between observed and counterfactual potential outcome distributions. The characterization highlights the restrictiveness of RS, motivating a weaker identifying condition. Under this alternative, we construct identifying bounds on the distributional treatment effects of interest through a linear semi-infinite programming (SILP) formulation. Our identification strategy also clarifies how richer exogenous instrument variation, such as multi-valued or multiple instruments, can further tighten these bounds. Finally, exploiting the SILP's saddle-point structure and Karush-Kuhn-Tucker (KKT) conditions, we establish large-sample properties for the empirical SILP: consistency and asymptotic distribution results for the estimated bounds and associated solutions.


In the treatment effect literature, RS (or in its simplest form \textit{Rank Invariance}) is widely used and plays a central role in identifying treatment effects in the presence of self-selection into treatment; see, e.g., \citet{foresi1995conditional}, \citet{heckman1997making}, \citet{chesher2003identification,Che05}, \citet{chernozhukov2005iv}, \citet{VY07}, \citet{jun2011tighter}, \citet{SV11}, \citet{d2015identification}, \citet{torgovitsky2015identification}, \citet{vuong2017counterfactual}, \citet{han2021identification}, among many others. Specifically, RS requires that, within any given compliance group, the population ranks of the treated and untreated potential outcomes have the same distribution. The plausibility of this assumption is often questionable in applications and lacks firm theoretical justification (see, e.g., \citet{maasoumi2019gender}, and also \citet{ChernozhukovHansen2013ARE} for a survey). In response, several papers propose tests for the validity of RS; see \citet{frandsen2018testing}, \citet{dong2018testing}, and \citet{kim2022testing}.

In this paper, we characterize RS by the model restrictions it imposes on the relationship between observed and counterfactual outcome distributions. Under a mild additional condition, we show that RS is equivalent to a two-sided preservation of first-order stochastic dominance (FOSD) across treatment statuses: for any two groups formed as linear mixtures  of compliance types, if one group's treated-outcome distribution is FOSD-dominated by the other's, then the same dominance holds for their untreated counterfactual-outcome distributions; and vice versa. This connection between RS and FOSD preservation across potential-outcome distributions appears to have received limited attention in the literature. By establishing this equivalence, we bridge the structural IV formulation of \citet{chernozhukov2005iv} with the potential-outcomes framework of \citet{rubin1974estimating}.

Furthermore, we propose a relaxation of RS by introducing a weaker condition based on one-way preservation of FOSD. This new condition can be interpreted as requiring that the potential outcome without treatment is \textit{Rank Noisier} (RN) than the treated potential outcome. Our proposed RN notion  is related to, but distinct from, the \textit{Noisier} concept developed in \citet{pomatto2020stochastic}. We further show that this condition arises naturally in a broad class of structural models derived from primitive economic assumptions and motivated by empirical applications. Under this weaker condition, we derive bounds on distributional treatment effects via an SILP formulation.

Nonparametric identification of treatment effects with essential endogeneity has long been a challenging goal. For instance, \citet{manski1990nonparametric, manski1997monotone, MP00}, among others, construct sharp bounds on the average treatment effect (ATE) under assumptions on the direction of treatment effects and on selection, while allowing instruments to be invalid in specific ways. Even with valid instruments, however, bounds on the ATE are typically wide and uninformative for precise policy prediction. The local ATE (LATE) \citep[][]{imbens1994identification} and local QTE \citep{abadie2002instrumental} have been popular alternatives when we impose a monotonicity assumption on selection to treatment. However, the local group for which the treatment effect is identified may not be the group of policy interest. Therefore, the extrapolation of the local parameters is crucial for policy analysis (e.g., treatment allocation), in which case the identification challenge still remains \citep[see e.g.][]{mogstad2018using,han2020sharp}.

Building on our identification strategy, we further solve empirical SILP problems to construct estimators for our bounds and establish their large-sample properties. We reformulate the SILP as a saddle-point problem \citep[see e.g.][]{ChristensenConnault2023}. Following, e.g., \citet{shapiro1991asymptotic,ekeland1999convex}, we show that the optimal-value map is Hadamard directionally differentiable with respect to the function-valued SILP coefficients to be estimated from the data. This establishes consistency and an asymptotic distribution for the bound estimators; See also \citet{GoffMbakop2025} for related  results in finite-dimensional linear programming problems.

In our SILP, the optimal-value functional is, in general, only Hadamard directionally (rather than fully) differentiable with respect to the SILP coefficients. As emphasized by \citet{fang2019inference}, this typically leads to non-Gaussian limiting distributions and can invalidate standard bootstrap procedures for inference. To obtain the full differentiability, we impose a local stability condition requiring uniqueness of the primal-dual optimal pair \citep[see e.g.][for uniqueness implying smoothness and regularity]{bonnans2000perturbation}.   Under this condition,   the set of binding constraints forms a smooth manifold near the optimum, ensuring that both the primal and dual solutions vary smoothly under small perturbations of the SILP coefficients.

In addition, we exploit the SILP's saddle-point structure and KKT conditions to establish large sample properties for the optimal solution. Without assuming a full rank condition on the set of active constraints, we restructure the problem as a two-step nested optimization: an inner SILP that solves for a subvector of the solution while holding the other components fixed, and an outer problem that minimizes a convex criterion given by the inner problem's value function. Applying \citet{milgrom2002envelope}'s generalized envelope theorem, we obtain first- and second-order conditions for the outer problem, which we use to derive the asymptotic distribution for its solution.  Moreover, exploiting the KKT conditions for the inner SILP, we obtain the limiting distribution of the remaining components. A key insight to this derivation is to combine complementary slackness with primal feasibility, providing us first-order conditions from the slackness equalities, since the slack attains its local minimum (or maximum) value at the active constraints. Our analysis contributes to the growing literature on inference for optimization-based estimators that confronts nondifferentiability and nonstandard asymptotics; see, e.g., \citet{andrews1999estimation,chernozhukov2007estimation,fang2019inference,hsieh2022inference,ChristensenConnault2023,fang2023inference,GoffMbakop2025}, among many others.

The paper is organized as follows. Section~\ref{sec model} introduces the treatment effect model, along with the necessary notation and assumptions. To motivate our approach, we begin by characterizing the RS assumption and then introduce a new condition that relaxes RS. We then derive bounds for the distributional treatment effects of interest under this weaker condition. Section~\ref{sec:Systematic-Calculation-of} discusses the computation of these bounds by solving an  SILP problem, and Section~\ref{sec asy} establishes the asymptotic properties of the estimated bounds derived from the empirical SILP problem. Section~\ref{sec:Numerical-Studies} presents numerical studies to illustrate our method. All proofs are provided in Appendix~\ref{sec:Proofs}.

\section{The Model and Bounds on Treatment Effects\label{sec model}}

This section presents a treatment effect model that accounts for individual self-selection  into treatment, while  allowing treatment effects to vary across individuals with identical observed characteristics (i.e. covariates). 
To fix ideas, and also in line with both the econometrics and statistics literature, we adopt Rubin's potential outcome framework \citep[see, e.g.,][]{heckman1986alternative,imbens1994identification}, 
augmented  with structural assumptions that address treatment selection.  We then motivate our key identification assumption and use it to derive bounds on the  distributional treatment effects of interest.

\subsection{Framework}
Let $Y_1$ denote the counterfactual outcome under treatment and $Y_0$ the counterfactual outcome without treatment. For simplicity, we assume both $Y_1$ and $Y_0$ are continuously  distributed. The observed outcome $Y \in \mathcal{Y} \subseteq \mathbb{R}$ is defined as $Y \equiv Y_D$, where $D \in \{0, 1\}$ is the observed treatment indicator, representing an individual's  treatment decision in response to instrumental variables (IVs) $Z$ and other covariates $X \in\mathcal X\subseteq  \mathbb{R}^{d_X}$. We assume that $Z$ is either a vector of binary IVs or a multi-valued IV, taking $L$ distinct values, i.e., $Z \in \mathcal{Z} \equiv \{z_1, \dots, z_L\}$. Multi-valued or multiple IVs are common in both observational studies (e.g., multiple natural experiments affecting the same policy) and experimental studies (e.g., randomized control trials with multiple treatment arms implemented simultaneously or sequentially).\footnote{See \citet{mogstad2021causal} for a recent survey.}

Following \citet{chernozhukov2005iv}, each latent outcome $Y_d$ is expressed in terms of its quantile function. Specifically, for  $d \in \{0,1\}$,  
\[
Y_d = q(d, X, U_d),
\]  
where $q(d, x, \cdot)$ is  the {\it quantile function} of $Y_d$, given $X=x$. Because $Y_d$ is continuously distributed, $U_d \sim U(0,1)$ and $q(d,x,\cdot)$ is strictly monotone. Note that $U_d$ represents the \textit{rank} of $Y_d$ within its population cumulative distribution, i.e., $U_d=F_{Y_d|X}(Y_d|X)$.  Moreover, the treatment selection $D$ is given by
\[
D = h(Z, X, \eta),
\]
where function $h$ describes treatment selection by the individual,  and $\eta\in\mathcal T\subseteq \mathbb R^{d_\eta}$ is a random vector capturing  unobserved heterogeneity.  As noted by \citet{imbens1994identification}, individuals can be classified into distinct compliance types based on how their treatment choices would  respond to each possible value of  the IVs, as determined by $\eta$ (with covariates $X$ fixed).

In our analysis, we focus on evaluating the treatment effects on the distributional features of potential outcomes $Y_d$ for the treated (or untreated) populations. For $d\in\{0,1\}$
and $x\in\mathcal{X}$, the {\it quantile treatment effects} (QTE) at  $\tau\in(0,1)$ for the group with treatment status $D=d$ and covariates $X=x$ is define as: 
\begin{align*}
QTE_{\tau}(d,x) =Q_{Y_{1}|D,X}(\tau| d,x)-Q_{Y_{0}|D,X}(\tau| d,x),
\end{align*}
for $\tau\in(0,1)$.\footnote{$Q_{A| B}(\tau| b)\equiv \inf\{a\in\mathbb{R}: F_{A| B}(a| b)\ge \tau\}$ for (generic) random variable $A$ and random element $B$.} Similarly, the {\it (conditional) average treatment effects} (ATE) for the population with the treatment status $D=d$ and $X=x$ is given by:
\begin{align*}
ATE(d,x) & =E(Y_{1}-Y_{0}| D=d,X=x).
\end{align*} Note that the unconditional QTE/ATE can be derived when these parameters are identified across all $d\in\{0,1\}$ and $x\in\mathcal{X}$. These parameters are of significant interest to researchers and policymakers for assessing the impact of interventions in policy evaluations. Because our model is fully nonparametric,  we fix $X=x$ throughout this section.

Throughout, we maintain the  assumption that the IVs are valid, i.e.,  the instrument $Z$ must be (strongly) relevant to individuals' treatment selection, and satisfy the exclusion restriction. 
\begin{myas}{Z}\label{as:Z}
For $d\in\{0,1\}$, $Z\perp (Y_{d},\eta)\, | X$. Moreover, $E(D|X,Z)\neq E(D|X)$ almost surely. 
\end{myas}
\noindent Assumption \ref{as:Z} has  been widely assumed in the literature; see e.g.  \citet{imbens1994identification}.  Applying $Y_d$'s quantile expression, the first half of Assumption \ref{as:Z} can be equivalently rewritten as $Z\perp (U_{d},\eta)\, | X$. In Assumption \ref{as:Z},  note that covariates $X$ are not required to be exogenous with respect to either the potential outcomes or the IVs.

\subsection{Motivations and Characterization of the RS Condition}

To identify treatment effects under nonseparable endogeneity, a prevalent approach in the literature is to restrict heterogeneity in treatment effects via RS (or, in its simplest form, rank invariance).

\begin{definition}[Rank Similarity, RS]  
Fix $X = x$. We say that potential outcomes $Y_0$ and $Y_1$ have similar ranks if their rank variables $U_0$ and $U_1$ are identically distributed conditional on $\eta$, i.e.,  
\[
F_{U_0 |X, \eta}(\cdot | x, t) = F_{U_1 | X, \eta}(\cdot | x, t), \ \ \forall \ t \in \mathcal T.
\]  


\end{definition}

By definition, RS ensures that the rank distribution of $Y_d$   remains unchanged across treatment status $d\in\{0,1\}$, regardless of the compliance type. As noted by \citet{chernozhukov2005iv}, the strongest yet simplest form of the RS condition is \textit{Rank Invariance}, which requires that $U_0=U_1$ almost surely. 



Interestingly, we notice that RS is inherently linked to  the preservation of first-order stochastic dominance (FOSD)  between  potential outcome distributions to their counterfactual counterparts, a connection that appears to have received limited attention in the existing literature.

\begin{mycondi}{S}\label{condi_2}Fix $x\in\mathcal{X}$.
For any absolutely continuous  function $G:\mathcal T\rightarrow \mathbb R$ and  constant $c\in\mathbb R_+$,  the following inequality:
\[
\int_\mathcal T F_{Y_{1}|X, \eta}(\cdot\, | x,t)dG(t)\leq c,
\]
holds if and only if 
\[
\int_\mathcal T F_{Y_{0}|X, \eta}(\cdot\, | x,t)dG(t)\leq c.
\]
\end{mycondi}

Condition \ref{condi_2} describes a biconditional relationship for linear mixtures of potential outcome distributions and their counterfactual counterparts, particularly ensuring the preservation of FOSD between potential outcome distributions. To see this, let \( W(\cdot) \) and \( \tilde{W}(\cdot) \) be arbitrary two probability distribution functions defined over \( \mathcal{T} \). Then, Condition \ref{condi_2} implies that: the following FOSD relationship
\[
\int_\mathcal T F_{Y_{1}|X,\eta}(\cdot|x,t)dW(t)\leq \int_\mathcal T F_{Y_{1}|X,\eta}(\cdot|x,t)d\tilde W(t)
\]
holds if and only if 
\[
\int_\mathcal T F_{Y_{0}|X,\eta}(\cdot|x,t)dW(t)\leq \int_\mathcal T F_{Y_{0}|X,\eta}(\cdot|x,t)d\tilde W(t).
\]This result can be derived by defining \( G(\cdot) = W(\cdot) - \tilde{W}(\cdot) \) and setting \( c = 0 \) in Condition~\ref{condi_2}. 

The following lemma shows that under an additional ``regularity'' condition, RS can be equivalently characterized by Condition~\ref{condi_2}. 

\begin{lemma}
\label{lemma1}  
Fix \( X = x \). Assume that the function \( q(d, x, \cdot) \) is strictly monotone. Then, RS implies Condition \ref{condi_2}. Additionally, suppose that for any \( u \in (0,1)\), there exists an absolutely continuous  function \( W_u: \mathcal{T} \to \mathbb{R} \) such that
\[
\int_{\mathcal T} F_{U_1|X,\eta}(\cdot| x,t) \, dW_u(t) = \mathbf{1}(\cdot \geq  {u}),
\]where $\mathbf 1(\cdot)$ is the indicator function, and moreover, 
\[
\int_{\mathcal T} F_{U_0|X,\eta}(\cdot| x,t) \, dW_u(t) = \int_{\mathcal T} F_{U_1|X,\eta}(\cdot| x,t) \, dW_u(t).
\]  Then
RS holds if Condition \ref{condi_2} is satisfied.
\end{lemma}
\proof see Appendix \ref{proof_lemma1}.\qed

Lemma~\ref{lemma1} shows that Condition~\ref{condi_2} is a key component of RS. To see this, note that RS can be equivalently represented as follows: For any absolutely continuous function \(G:\mathcal T\to\mathbb R\),
\[
\int_{\mathcal T} F_{U_0| X,\eta}(\cdot| x,t)\,dG(t)
=
\int_{\mathcal T} F_{U_1| X,\eta}(\cdot| x,t)\,dG(t).\footnote{A simple sketch of the proof: If RS holds, integrating both sides with respect to $dG$ gives this representation. On the other hand, restricting $G$ to point-mass functions recovers the standard RS formula.}
\]
Thus, under Condition~\ref{condi_2}, Lemma~\ref{lemma1} implies that the RS equivalence extends from the class \(\{W_u(\cdot): u\in(0,1)\}\) to all absolutely continuous functions \(G\).

\subsection{Introducing Key Conditions}

In this subsection, we propose two  alternative assumptions to relax RS. The first one corresponds to the ``only if'' part of Condition \ref{condi_2}, while the other corresponds to its ``if'' part. It is worth pointing out that we also drop the regularity condition in Lemma~\ref{lemma1} for our identification analysis.  Grounded in economic theory, the proposed two assumptions have distinct empirical applicability; depending on the context, researchers may find that one, or both, applies.

\begin{mycondi}{S$_{1}$}\label{as:condi_1}
Fix $x\in\mathcal{X}$. For any absolutely continuous  function $G:\mathcal T\rightarrow \mathbb R$ and  constant $c\in\mathbb R_+$, the following condition holds: if
\[
\int_\mathcal T F_{Y_{1}|X, \eta}(\cdot| x,t)dG(t)\leq c \quad \Longrightarrow
\quad \int_\mathcal T F_{Y_{0}|X, \eta}(\cdot| x,t)dG(t)\leq c.
\]
\end{mycondi}
\begin{remark}
Condition~\ref{as:condi_1} is the ``only-if'' part of Condition~\ref{condi_2} and is therefore  weaker than RS.
As we will discuss in details later, \ref{as:condi_1} holds when the distribution of \(Y_0\) is obtained from that of \(Y_1\) by a monotone mixture (i.e., \(Y_0\) is {\bf{rank noisier}} than  \(Y_1\)). 
Formally, there exist a random variable \(\xi\) with CDF \(F_\xi\)  and a family of nondecreasing functions \(\{\psi_x(\cdot\,,s): s\in\mathcal S_\xi\}\) such that, 
\[
F_{Y_{0}\,| X,\eta}(\cdot\,| x,t)=
\int F_{Y_{1}| X,\eta}\!\big(\psi_x(\cdot,  s)\,| x,t\big)\, dF_\xi(s)
\]
holds for  all  \(t\in\mathcal T\). Clearly, Condition~\ref{as:condi_1} generalizes the counterfactual mapping  in \citet{vuong2017counterfactual}. 
\end{remark}
\begin{remark}
It is worth pointing out an alternative to Condition~\ref{as:condi_1},  slightly weaker but of the same flavor. 
Let $W$ and $\tilde W$ be two probability measures on $\mathcal T$. If the following inequality
\[
\int_{\mathcal T} F_{Y_{1}| X,\eta}(\cdot\, | x,t)\, dW(t) \;\leq\; \int_{\mathcal T} F_{Y_{1}| X,\eta}(\cdot\,| x,t)\, d\tilde W(t)
\]hold,  
then 
\[ 
\int_{\mathcal T} F_{Y_{0}| X,\eta}(\cdot\,| x, t)\, dW(t) \;\le\; \int_{\mathcal T} F_{Y_{0}| X,\eta}(\cdot\,| x, t)\, d\tilde W(t).
 \]
This alternative condition can be interpreted as a one-way preservation of FOSD for the potential-outcome distributions across two mixtures of the population
\end{remark}

Under Assumption \ref{as:Z}, we can further derive explicit model restrictions implied by Condition \ref{as:condi_1}. For any $(x, z) \in \mathcal{X} \times \mathcal{Z}$ and $d \in \{0,1\}$, note that
\[
\Pr(Y_d \leq \cdot, D = 1 | Z = z, X = x) 
= \int_{\mathcal{T}} \Pr(Y_d \leq \cdot, D = 1 | \eta = t, Z = z, X = x) \, dF_{\eta | X,Z}(t | x, z).
\]
Since $D$ is binary and satisfies $D = h(Z, X, \eta)$, we have
\[
\Pr(Y_d \leq \cdot, D = 1 | \eta = t, Z = z, X = x) = \Pr(Y_d \leq \cdot | \eta = t, Z = z, X = x)\times h(z,x,t),
\]where $h(z,x,t)$ takes the binary value $0$ or $1$. 
By Assumption \ref{as:Z}, it follows that
\[
\Pr(Y_d \leq \cdot, D = 1 | Z = z, X = x) = \int_{\mathcal{T}} F_{Y_d | X,\eta}(\cdot | x, t) \, dG(t |x, z),
\]
where $G(t |x, z) \equiv \int_{-\infty}^{t} h(z, x, u) \, dF_{\eta | X}(u | x)$. Thus, Condition \ref{as:condi_1} implies the following model restriction: For any $\gamma_0\in\mathbb R$ and $ (\gamma_{11}, \dots, \gamma_{1L}) \in \mathbb{R}^{L}$, if
\begin{align}
\label{eq:2.1}
\sum_{\ell=1}^{L} \gamma_{1\ell} \Pr(Y_{1} \leq \cdot, D = 1 | Z = z_{\ell}, X = x) \leq \gamma_0,
\end{align}
then
\begin{align}
\label{eq:2.2}
\sum_{\ell=1}^{L} \gamma_{1\ell} \Pr(Y_{0} \leq \cdot, D = 1 | Z = z_{\ell}, X = x) \leq \gamma_0.
\end{align}
As will be discussed later, the above condition will be crucial for deriving bounds on distributional treatment effects.

Similarly, we define the converse of Condition \ref{as:condi_1}, denoted as Condition \ref{as:condi_0}. 
\begin{mycondi}{S$_{0}$}\label{as:condi_0}
Fix $x\in\mathcal{X}$. For any absolutely continuous function $G: \mathcal{T} \rightarrow \mathbb{R}$ and constant $c \in \mathbb{R}_+$, if 
\[ 
\int_{\mathcal{T}} F_{Y_{0} | X, \eta}(\cdot | x,t) \, dG(t) \leq c\quad \Longrightarrow
\quad \int_{\mathcal{T}} F_{Y_{1} | X, \eta}(\cdot | x,t) \, dG(t) \leq c. 
\] 
\end{mycondi}
Clearly, combining Conditions \ref{as:condi_1} and \ref{as:condi_0} together are equivalent to Condition \ref{condi_2}.  It is of interest to understand when Condition \ref{as:condi_1}, Condition \ref{as:condi_0}, or both would hold in practice. To illustrate, we now derive e.g. Condition \ref{as:condi_1} from primitive conditions on the joint distribution of $(U_0, U_1)$.

Following \citet{vuong2017counterfactual}'s counterfactual mapping approach,  consider the mapping from \( F_{U_1| X, \eta}(\cdot | x, t) \) to \( F_{U_0 | X, \eta}(\cdot | x, t) \) for each \( t \in \mathcal{T} \). By  the law of iterated expectations, we have
\[
F_{U_0 | X, \eta}(\cdot\, | x, t) = \int_0^1 F_{U_0 | U_1, X, \eta}(\cdot \,| u_1, x, t) \, dF_{U_1 | X, \eta}(u_1 | x, t).
\]
This  indicates that  there exists a linear operator that maps the distribution \( F_{U_1| X, \eta}(\cdot \,| x, t) \) to \( F_{U_0 | X, \eta}(\cdot\, | x, t) \), with the conditional distribution \( F_{U_0 | U_1, X, \eta}(\cdot\, | \cdot, x, t) \) serving as the kernel function. Suppose in addition that \( U_0 \perp \eta \, | U_1, X \). Then, for any absolutely continuous function $G: \mathcal{T} \rightarrow \mathbb{R}$, 
\[
\int_\mathcal T F_{U_0 | X, \eta}(\cdot\, | x, t) \, d G(t)= \int_0^1 F_{U_0 | U_1, X}(\cdot\, | u_1, x) \, d\left\{\int_\mathcal TF_{U_1 | X, \eta}(u_1 | x, t)\, dG(t)\right\},
\]
which is a Fredholm integral equation of the first kind. In this literature, significant attention has been given to understanding how the function \( \int_\mathcal T F_{U_0 | X, \eta}(\cdot | x, t) \, d G(t) \) can inform us about the behavior of its inverse function \( \int_{\mathcal T}F_{U_1 | X\eta}(\cdot |x, t)\, dG(t) \) under certain conditions on the kernel function $F_{U_0 | U_1, X}(\cdot |\cdot, x)$; See e.g. \cite{buchholz2021semiparametric}.




\begin{myas}{I}\label{as:I}
Assume that \( U_0 \) is independent of \( \eta \) conditional on \( U_1 \) and \( X \), i.e.,
\[
U_0 \perp \eta \, | U_1, X.
\]
Moreover, assume that \( U_0 \) is positively regression dependent on \( U_1 \); that is,  for  any fixed $x$ and \( u_0 \in (0, 1) \), the conditional distribution  \( F_{U_0 | U_1,X}(u_0 | u_1, x) \) is non-increasing in \( u_1 \in (0, 1) \).
\end{myas}
\noindent
In Assumption \ref{as:I}, the first part implies that the rank \( U_1 \) captures all the information related to the compliance type.  The second part introduces a weaker notion of positive dependence, i.e., positive regression dependence, than commonly used alternatives such as positive affiliation or decreasing inverse hazard rate \citep[see, e.g.,][]{Castro2007}. In the following discussion, we will present structural examples in which Assumption \ref{as:I} is satisfied.


Let \( \mathcal{L}^1_+([0,1]) \) denote the set of all non-negative Lebesgue integrable functions on the interval \([0,1]\), i.e.,
\[
\mathcal{L}^1_+([0,1]) = \left\{ g: [0,1] \to \mathbb{R}_+ \,\Big|\,  \int_0^1 g(t)\, dt < \infty \right\}.
\]
Let further
    \[
    \mathcal{W}(x) \equiv \left\{ -\frac{\partial F_{U_0 | U_1, X}(u_0 \,| u_1, x)}{\partial u_1} : u_0 \in [0,1] \right\} 
    \] be the collection of functions in $u_1\in[0,1]$, derived from $F_{U_0 | U_1, X}(u_0 \,| u_1, x)$.

\begin{lemma}
\label{lemma:2.2}
Fix $x\in\mathcal X$. Suppose that the function \( q(d, x, \cdot) \) is strictly monotone and that Assumption \ref{as:I} holds. Then Condition \ref{as:condi_1} is satisfied. In addition, suppose that (i) $F_{U_0| U_1, X}(u_0 | 1, x) = 0 $ and $F_{U_0| U_1, X}(u_0 | 0, x) = 1 $ hold for any $u_0\in(0,1)$; (ii)
 the linear span of \( \mathcal{W}(x) \) is dense in \( \mathcal{L}^1_+([0,1]) \).
Then Condition \ref{as:condi_0} holds.
\proof see Appendix~\ref{proof_lemma2.2}.\qed
\end{lemma}



In Lemma~\ref{lemma:2.2}, the first part shows that Condition~\ref{as:condi_1} follows directly from Assumption~\ref{as:I}. This suggests tha Condition~\ref{as:condi_1}  can be justified on economic grounds. Similarly, one can consider the counterpart to Assumption \ref{as:I} by switching the roles of \( U_0 \) and \( U_1 \), so that Condition \ref{as:condi_0} holds under an alternative economic interpretation. The second part of the lemma is a topological condition. Specifically, Condition (ii) requires that the collection \( \mathcal{W}(x) \), derived from the kernel function \( F_{U_0 | U_1, X}(\cdot \,| \cdot, x) \), is sufficiently rich to approximate any function in \( \mathcal{L}^1_+([0,1]) \) through linear combinations. This condition can be satisfied in some settings, e.g., \( F_{U_0 | U_1, X}(\cdot| u_1, x) = \mathbf{1}(\cdot \leq u_1) \) for any $u_1\in[0,1]$. In general, however, it is difficult to motivate for this condition or  to verify it empirically.

%

\subsection{Bounds on Distributional Treatment Effects}
Given Condition \ref{as:condi_1} or Condition \ref{as:condi_0}, we are now prepared to construct bounds for distributional treatment effects in policy analysis. In the following discussion, we maintain Condition \ref{as:condi_1} (respectively, Condition \ref{as:condi_0})  and consider the quantile treatment effect \( QTE_{\tau}(1,x) \) (respectively, \( QTE_{\tau}(0,x) \)) at  quantile \( \tau \in (0,1) \). By definition, note that
\[
QTE_{\tau}(1, x) = Q_{Y_1 | D, X}(\tau | 1, x) - Q_{Y_0 | D, X}(\tau | 1, x),
\]
where \( Q_{Y_1 | D, X}(\tau | 1, x) \) is trivially identified by the ``observed'' conditional quantile \( Q_{Y | D, X}(\tau | 1, x) \). Thus, identifying \( QTE_{\tau}(1, x) \) reduces to identifying the counterfactual quantile  \( Q_{Y_0 | D, X}(\cdot\, | 1, x) \), which requires to invert the conditional distribution \( F_{Y_0 | D, X}(\cdot \,| 1, x) \). For simplicity, we focus on constructing the lower bound of \( Q_{Y_0 | D, X}(\tau | 1, x) \) in the following discussion. It is straightforward to extend our approach analogously to the upper bound.



To proceed, we  apply Condition \eqref{eq:2.1} - \eqref{eq:2.2} to a particular class of weights \( (\gamma_0, \gamma_{11}, \ldots, \gamma_{1L}) \in \mathbb{R}^{L+1} \) satisfying \( \sum_{\ell=1}^L \gamma_{1\ell} = 0 \), and then obtain the following result.

\begin{theorem}\label{theorem2}
Suppose Assumption \ref{as:Z} and Condition \ref{as:condi_1} hold. For $\gamma_0\in\mathbb R$, and \( (\gamma_{11},\cdots,\gamma_{1L}) \in \mathbb{R}^{L} \) satisfying \( \sum_{\ell=1}^L \gamma_{1\ell} = 0 \), assume that
\begin{equation}
\label{eq2.3}
\Pr(Y \leq \cdot \, | D = 1, X = x) \leq \gamma_0 + \sum_{\ell=1}^L \gamma_{1\ell}\, \Pr(Y \leq \cdot, D = 1 | Z = z_\ell, X = x).
\end{equation}
Then the counterfactual distribution \( F_{Y_0 | D, X}(\cdot | 1, x) \) is bounded above as follows:
\begin{equation}
\label{eq2.4}
\Pr(Y_0 \leq \cdot | D = 1, X = x) \leq \gamma_0 - \sum_{\ell=1}^L \gamma_{1\ell}\, \Pr(Y \leq \cdot, D = 0 | Z = z_\ell, X = x).
\end{equation}
\proof see Appendix~\ref{proof_theorem2}.\qed
\end{theorem}

Because there may exist multiple (indeed, infinitely many) vectors \( (\gamma_0, \gamma_{11}, \ldots, \gamma_{1L}) \) satisfying \( \sum_{\ell=1}^L \gamma_\ell = 0 \) and eq. \eqref{eq2.3}, we aim to further tighten the bounds. To this end, we first introduce some notation to simplify the exposition. For each \( (z, x) \in \mathcal{Z} \times \mathcal{X} \), let \( p(z, x) = \Pr(D = 1 | Z = z, X = x) \) be the propensity score. Moreover, for \( d \in \{0,1\} \), define
\[
\Delta_{dz}(\cdot \, | x) \equiv \Pr(Y \leq \cdot, D = d \, | Z = z, X = x) 
- \Pr(Y \leq \cdot, D = d \, | Z = z_L, X = x),
\]
supported on \( [\underline{y}, \overline{y}] \). In the above definition, we treat $z_L$ as the reference point of $Z$ on the support $\mathcal Z$.  Next, define the \((L-1)\)-dimensional profile vector of functions
\[
\Delta_d(\cdot \, | x) \equiv \big( \Delta_{dz_1}(\cdot \, | x), \ldots, \Delta_{dz_{L-1}}(\cdot \, | x) \big) \in \mathbb{R}^{L-1},
\]
which captures the variation in \( \Delta_{dz}(\cdot | x) \)  induced by the instrument.   Finally, we denote \( \gamma_1 \equiv (\gamma_{11}, \ldots, \gamma_{1,L-1})' \in \mathbb{R}^{L-1} \) as a column vector. Note that by letting \( \gamma_{1L} = -\sum_{\ell=1}^{L-1} \gamma_{1\ell} \), the constraint \( \sum_{\ell=1}^L \gamma_{1\ell} = 0 \) is automatically satisfied.


%

\begin{corollary}\label{corollary1}
Fix \( X = x \), and suppose Assumption \ref{as:Z} and Condition \ref{as:condi_1} hold. Then we obtain an upper bound on \( F_{Y_0 | D, X}(\cdot \,| 1, x) \) as follows: for each \( y_0 \in \mathcal{Y} \),
\begin{align*}
F^{UB}_{Y_0 | D, X}(y_0 | 1, x) &\equiv \min_{(\gamma_0,\gamma_1') \in\mathbb R^L}\quad \gamma_0-\gamma_1' \Delta_0(y_0 | x) \\
\text{s.t.}  & \quad F_{Y | D, X}(y | 1, x) \leq \gamma_0+\gamma_1' \Delta_1(y | x), \quad \forall \, y \in [\underline{y},\, \overline{y}]. \nonumber
\end{align*}
\end{corollary}

%

\noindent
The proof of Corollary~\ref{corollary1} is straightforward and thus omitted. Together, Theorem~\ref{theorem2} and Corollary~\ref{corollary1} highlight the identifying power of multi-valued  IVs. In particular, in the above LP problem for the upper bound \( F^{UB}_{Y_0 | D, X}(\cdot | 1, x)\), its feasible region depends critically on the variations of the  IVs.

To see this, consider the linear subspace of a Hilbert space (e.g., \( L^2([\underline{y}, \, \overline{y}]) \)) spanned by the set of functions \( \{1\} \cup \{ \Delta_{dz_\ell}(\cdot | x) : \ell \leq L \} \). Suppose further that each \( \Delta_{dz_\ell}(\cdot | x) \) belongs to the logistic family of distribution functions, and let \( L \to \infty \). Then the resulting linear subspace becomes dense in the space of distribution functions on \( [\underline{y}, \overline{y}] \), allowing for arbitrarily accurate approximation of $F_{Y | D, X}(y | 1, x) $; See, e.g., \citet{hornik1989multilayer}. In this sense, \( F_{Y_0 | D, X}(\cdot | 1, x) \) becomes point identified in the limit by \( F^{UB}_{Y_0 | D, X}(\cdot | 1, x) \). On the other hand, when \( L \) is small, even an extra variation of $\mathcal Z$ expands the dimensionality of the linear subspace spanned by \( \{1\} \cup \{ \Delta_{dz_\ell}(\cdot | x) : \ell \leq L \} \), potentially tightening the bounds on \( F_{Y_0 | D, X}(\cdot | 1, x) \) substantially. In Section~\ref{sec:Numerical-Studies}, we investigate identification power of  the IVs' variation by using Monte Carlo studies.


Finally, we construct a lower bound for $QTE_{\tau}(1,x)$ from \( F^{UB}_{Y_0|D,X}(\cdot\, | 1, x) \). Using the worst case bounds for the conditional quantile
\citep[see e.g.][]{manski1994selection,blundell2007changes}, we have 
\begin{align*}
 Q_{Y_{0}|D,X}(\tau | 1,x)\ge Q_{Y_{0}|D,X}^{LB}(\tau | 1,x),
\end{align*}
where $Q_{Y_{0}|D,X}^{LB}(\tau|1,x)$ is the $\tau$-th quantile of $F_{Y_{0}|D,X}^{UB}(\cdot\, |1,x)$.  Because $F_{Y_{0}|D,X}^{UB}(\cdot\, |1,x)$ is point-wisely constructed, it may not be  monotone over its support. Next, we define $Q^{LB}_{Y_{0}|DX}(\tau|1,x)$ by inverting the upper bound function $F_{Y_{0}|D,X}^{UB}( \cdot\, |1,x)$ as follows:
\[
Q_{Y_{0}|D,X}^{LB}(\tau|1,x)= \inf_{ y\in\mathcal Y}\, \left\{ y: F_{Y_{0}|D,X}^{UB}( y\, |1,x)\geq \tau\right\}.
\]
Furthermore, we construct an upper bound for $QTE_{\tau}(1, x)$ by
\[
QTE^{UB}_{\tau}(1, x) =Q_{Y_1 | D, X}(\tau | 1, x) - Q^{LB}_{Y_0 | D, X}(\tau | 1, x).
\]


\subsection{Illustrative Examples}

In this subsection, we first present an illustrative example to clarify the intuition behind our identification strategy. Next, we provide empirical justifications for the proposed key conditions, i.e., Conditions \ref{as:condi_1} or \ref{as:condi_0}, by deriving them from primitive structural assumptions.

We begin with a setting involving multiple-valued IVs within \citet{imbens1994identification}'s LATE framework. Throughout, we fix \( X = x \). Assume that \( h(z_{\ell}, x, \cdot) \leq h(z_{\ell+1}, x, \cdot) \) for \( \ell = 1, \ldots, L \), and let \( D_z \equiv h(z, x, \eta) \) denote the potential treatment status when \( Z = z \) and \( X = x \). This setup leads to the following monotone selection condition: 
\begin{align}
\label{as: monotone selection}
D_{z_{\ell+1}} \geq D_{z_{\ell}} \quad \text{a.s.}, \quad\text{for} \ \ \ell=1,\cdots,L-1.
\end{align}
Under this condition, the observed treated group \( \{D = 1\} \) comprises both always-takers (AT) and compliers (C). Define the always-takers (AT)  as \( \{ \eta \in \mathcal{T} : D_1 = \cdots = D_L = 1 \} \), i.e., individuals who always receive treatment regardless of the IV value. For \( 1 \leq \ell \leq L - 1 \), define the complier group induced by shifting the instrument from \( z_\ell \) to \( z_{\ell+1} \) as:
\[
(z_\ell, z_{\ell+1})_C \equiv \left\{ \eta \in \mathcal{T} : D_{z_j} = 0 \text{ if and only if } j \leq \ell \right\}.
\]
For instance, \( (z_1,z_2)_C \) denotes eager compliers, and \( (z_{L-1}, z_L)_C \) denotes reluctant compliers, following the terminology of \citet{mogstad2021causal}. Within this LATE framework, the following lemma restates Theorem \ref{theorem2} under the monotone selection assumption.

\begin{lemma}\label{lem:refutable}
Suppose Assumption \ref{as:Z} and the monotone selection condition \eqref{as: monotone selection} hold. Then, Condition \ref{as:condi_1} implies the following: For $\gamma_0\in\mathbb R$ and $(\gamma_{11},\cdots,\gamma_{1L})\in\mathbb R^L$,  if
\[
\Big( \sum_{k=1}^L \gamma_{1k} \Big) \Pr(Y_1 \leq \cdot \,, \eta \in \text{AT}) 
\leq \gamma_0 - \sum_{\ell=1}^{L-1} \Big( \sum_{k=\ell+1}^{L} \gamma_{1k} \Big) 
\Pr\big[ Y_1 \leq \cdot \,, \eta \in (z_\ell, z_{\ell+1})_C \big],
\]
then it follows that
\[
\Big( \sum_{k=1}^L \gamma_{1k} \Big) \Pr(Y_0 \leq \cdot \,, \eta \in \text{AT}) 
\leq \gamma_0 - \sum_{\ell=1}^{L-1} \Big( \sum_{k=\ell+1}^{L} \gamma_{1k} \Big) 
\Pr\big[ Y_0 \leq \cdot \,, \eta \in (z_\ell, z_{\ell+1})_C \big].
\]
\end{lemma}

In Lemma \ref{lem:refutable}, the inequalities directly follow Conditions \eqref{eq:2.1}-\eqref{eq:2.2} under the monotone selection assumption, so the proof is omitted. According to Lemma \ref{lem:refutable}, if \( \sum_{k=1}^L \gamma_{1k} = 0 \), then Condition \ref{as:condi_1} yields testable model implications, since \( \Pr[Y_d \leq \cdot \,, \eta \in (z_\ell, z_{\ell+1})_C] \) is identified for both \( d = 0 \) and \( d = 1 \); See \citet{imbens1997estimating}. An exception arises when \( L = 2 \), as the condition holds trivially. When \( \sum_{k=1}^L \gamma_{1k} \neq 0 \), the model yields bounds on the counterfactual distribution \( \Pr(Y_0 \leq \cdot \,; \eta \in \text{AT}) \), and increasing \( L \) would tighten these bounds. Such extrapolation is central to  Theorem \ref{theorem2}.

Next, we examine Condition \ref{as:condi_1} under primitive structural assumptions. By Lemma \ref{lemma:2.2},  Condition \ref{as:condi_1} holds under Assumption \ref{as:I}, which has an intuitive interpretation: Rank $U_0$ is {\it noisier} than $U_1$.  
\begin{definition}[Rank Noisier, RN]  
Fix $X = x$. We say that rank of potential outcomes $Y_0$ is  noisier than that of $Y_1$ if their ranks $U_0$ and $U_1$ satisfy $ U_0=\psi_x(U_1,\xi)$ for some function $\psi_x$ strictly increasing in $U_1$, and random element $\xi\in\mathbb R^{d_\xi}$  independent of $(U_1,\eta)$ given $X=x$.
\end{definition}
Our RN concept extends the ``\textit{noisier}'' concept introduced by \cite{pomatto2020stochastic}, where a random variable \(Z'\) is considered noisier than \(Z\) if \(Z' = Z + W\) and \(W\) is independent of \(Z\). Hence, if $X + Z\prec_{FOSD} \tilde X + Z$ for some $Z$ that is independent of $X$ and $\tilde X$, then $X +Z' \prec_{FOSD}\tilde X + Z'$ for any independent $Z'$ that is {\it noisier} than $Z$. In contrast, our RN concept  supports a different type of preservation of stochastic dominance. Namely, for  absolutely continuous  function $G$ and $\tilde G$, if 
\[
\int_\mathcal T F_{Y_{1}|X,\eta}(\cdot\,| x,t)dG(t)\prec_{FOSD} \int_\mathcal T F_{Y_{1}|X,\eta}(\cdot\,| x,t)d\tilde G(t),
\]  then  
\[
\int_\mathcal T F_{Y_{0}|X,\eta}(\cdot\,| x,t)dG(t)\prec_{FOSD} \int_\mathcal T F_{Y_{0}|X,\eta}(\cdot\,| x,t)d\tilde G(t).
\] 
Note that the reverse  does not necessary hold for the same intuition provided in \cite{pomatto2020stochastic}.

We now illustrate that the  RN condition, namely, that \( Y_0 \) is rank-noisier than \( Y_1 \) or vice versa, can be justified within a class of empirical structural models.

\begin{example}[Auction]\label{ex1}
Consider an auction setting with both online and offline formats, where bids are generated according to
\[
B = \beta(D, X, V_D),
\]
with \( B \) denoting the bidder's submitted bid (which subsequently determines her revenue), \( D \in \{0,1\} \) indicating the auction format (\( D=1 \) for online and \( D=0 \) for offline), and \( X \) representing observed bidder or auction-specific covariates. The function \( \beta \) denotes the equilibrium bidding strategy, which depends on the auction format, the mechanism (e.g., first-price vs. second-price), and the bidder's characteristics and evaluation.

Moreover, let \(V_d = V + \xi_d\) denote the bidder's subjective valuation under format \(d\), where \(V\) is the objective valuation, and \(\xi_d\) is a format-specific idiosyncratic shock, assumed to satisfy \( \xi_d \perp (\eta, V)| X \). Online and offline formats may induce different bidding behaviors due to some behavior shocks. One may argue that $\xi_0$ is noisier  \citep[in the sense of][]{pomatto2020stochastic} than $\xi_1$, because, e.g.,  bidders in offline auctions are more emotionally influenced by the physical presence or behavior of other bidders, leading to more variable subjective valuations.
\end{example}

In Example~\ref{ex1}, the assumption that \( \xi_0 \) is noisier than \( \xi_1 \) implies that \( \xi_0 = \xi_1 + \zeta \) for some independent noise \( \zeta \). It follows that \( V_0 = V + \xi_0 = V_1 + \zeta \). Thus, conditional on \( X \), the rank of \( V_0 \) is more dispersed than the rank of \( V_1 \), consistent with the RN condition.  Importantly, this specific structure for the idiosyncratic shocks is not essential. For instance, one could instead impose   a nonlinear form as follows:
\[
\xi_0 = \max\{\phi(\xi_1), \zeta\},
\]
where \( \phi \) is a strictly increasing function and \( \zeta \) is again independent.

\begin{example}[Insurance/Vaccination]\label{ex2}
Suppose we are interested in evaluating the effect of insurance coverage or vaccination on health outcomes. Let \( Y \) denote a health outcome, and let \( D \in \{0,1\} \) indicate insurance or vaccination status, with \( D = 1 \) representing insured or vaccinated individuals.  Moreover, assume $Y=h(D,X,\epsilon_D)$, where  $X$ is a vector of observed covariates, \( \epsilon_d \) denotes the underlying treatment-specific health condition under treatment status \( d \), and assume that \( \epsilon_d \stackrel{d}{=} \epsilon + \xi_d \), where \( \epsilon \) captures individual's baseline health status, known to the individual (and thus potentially correlated with the decision \( D \)), and \( \xi_d \) represents unobserved health shocks specific to treatment status \( d \). 

In this setup, it is plausible that \( \xi_0 \) is noisier than \( \xi_1 \)   \citep[again, noisier in the sense of][]{pomatto2020stochastic}, as insurance or vaccination provides protection that helps prevent or manage predictable and well-understood health shocks.

\end{example}

In both Examples~\ref{ex1} and~\ref{ex2}, Condition~\ref{as:condi_1} holds under the assumption that \( U_0 \) is rank-noisier than \( U_1 \). In this case, Theorem~\ref{theorem2} and Corollary~\ref{corollary1} yield bounds on \( QTE_{\tau}(1, x) \), the quantile treatment effect for those who receive the treatment. The next example, however, illustrates the converse case.

\begin{example}[High-risk clinical trial]\label{ex3}
In contrast to Example~\ref{ex2}, suppose the treatment itself carries significant risk. Let \( D \in \{0,1\} \) indicate participation in a frontier medical trial, with \( D = 1 \) representing participation. In this case, it is plausible that $\xi_1$ is noisier than $\xi_0$ as newly developed treatments may involve substantial uncertainty due to unexpected side effects.
\end{example}
Example~\ref{ex3} provides justification for Condition~\ref{as:condi_0}, under which bounds on \( QTE_{\tau}(0, x) \), i.e., the quantile treatment effect for individuals who abstain from treatment, can be derived through a similar argument.  Assuming either Condition~\ref{as:condi_1} or~\ref{as:condi_0}, our approach yields partial identification of distributional treatment effects for the treated or the untreated, respectively.

In practice, it is often of interest to policymakers to determine which treatment parameter is most relevant for policy evaluation. Suppose the policymaker is primarily concerned with risk-averse individuals, who tend to prefer treatment options associated with lower uncertainty. For such a policymaker, she would be interested in evaluating the policy that  would offer a form of ``insurance'', i.e., either in the literal sense (e.g., health insurance) or through interventions that mitigate risk (e.g., vaccination, regulations on high-risk treatments). In this sense, our procedure offers a statistical tool for bounding the treatment effects for individuals with \( D = d \), when  rank \( U_{d} \) is less noisier than \( U_{1-d} \).



\section{Systematic Calculation of Bounds\label{sec:Systematic-Calculation-of}}

In this section, we employ optimization methods to systematically calculate the bounds introduced in Corollary~\ref{corollary1}. For clarity, we focus on the upper bound under the assumption that the IVs $Z$ are discrete, i.e., $Z \in \mathcal{Z} \equiv \{z_1, \dots, z_L\}$.  Extending the analysis to the case of continuous $Z$ is more challenging and left for future work. Since our model specification is fully nonparametric, we condition on $X = x$ throughout this section, regardless of whether $X$ is continuously or discretely distributed.

\subsection{Semi-Infinite LP and Regularity Conditions}
To begin with, we introduce some notation. For expositional simplicity, we assume that the domain $\mathcal{Y}$ is a finite interval on $\mathbb{R}$, i.e., $\mathcal{Y} = [\underline{y},\, \overline{y}]$.   For $d \in \{0,1\}$ and $z \in \mathcal{Z}$, recall that 
\[
\Delta_{dz}(\cdot| x) =\Pr(Y \leq \cdot, D = d | Z = z, X = x) 
- \Pr(Y \leq \cdot, D = d | Z = z_L, X = x)
\]
and
\[
\Delta_{d}(\cdot | x)=\big( \Delta_{dz_1}(\cdot| x),\cdots,\Delta_{dz_{L-1}}(\cdot| x)\big)
\] 
 supported on $[\underline y,\, \overline y]$.  By construction, $\Delta_{dz}(-\infty|x) = 0$ and $\Delta_{dz}(+\infty|z) = (-1)^{d+1}[p(z, x) - p(z_L, x)]$. Moreover,   $\Delta_{dz}(\cdot|x) $ is bounded and differentiable.

We now consider the LP problem introduced in Corollary~\ref{corollary1} for the upper bound of \( F_{Y_0 | D, X}(y_0|1, x) \), where \( y_0 \in [\underline{y}, \overline{y}] \). Let \( \mathbb{S}(x) \) denote the feasible region of the LP, defined as
\[
\mathbb{S}(x) = \left\{ (\gamma_0,\gamma'_1) \in \mathbb{R}^L : F_{Y | D, X}(y | 1, x) \leq \gamma_0+\gamma_1' \Delta_1(y | x) \quad \forall y \in [\underline{y}, \,\overline{y}] \right\}.
\]
By definition, \( \mathbb{S}(x) \) is constructed by a continuum of  inequality constraints, with \( F_{Y| D X}(\cdot | 1, x) \) and \( \Delta_1(\cdot | x) \)  as the intercept and slope functions, respectively. Moreover, it is straightforward that \( \mathbb{S}(x) \) is convex and unbounded. Thus, the upper-bound LP problem can  be rewritten as:
\[
F^{UB}_{Y_0 | D, X}(y_0 | 1, x) = \min_{(\gamma_0,\gamma'_1) \in \mathbb{S}(x)} \quad \gamma_0-\gamma_1' \Delta_0(y_0 | x).
\]
In this LP,  any feasible move in the direction \( \big(-1,\,\Delta_{0}(y_0| x)\big)\) decreases the objective value.

In the empirical LP problem introduced later, it is essential that the corresponding population LP satisfies certain regularity conditions such that the optimal value is stable with respect to small perturbations in these coefficients.


\begin{myas}
{C}\label{as:C}
For any $(z,x) \in \mathcal{Z} \times \mathcal{X}$, the conditional distribution function $F_{Y|DZX}(\cdot | 1, z, x)$ is absolutely continuous with respect to the Lebesgue measure on its support $[\underline{y}, \overline{y}]$. 
\end{myas}
\noindent
Under Assumption \ref{as:C}, the conditional distribution $F_{Y|DZX}(\cdot | 1, z, x)$ is continuous and differentiable a.e..  This assumption  ensures that  the functions $F_{Y|DX}(\cdot|1,x)$ and $\Delta_{1}(\cdot|x)$, which determine the LP constraints,  vary continuously over the compact support $[\underline{y}, \overline{y}]$.  



%
%

\begin{myas}{D}
\label{as:D}
Let $\mathbb{D}(x)$ denote the (unit-normalized) recession directions of the feasible set:
 \[
 \mathbb{D}(x) =\left\{(\delta_0,\delta_1')\in \mathbb R\times \mathbb R^{L-1}: \|(\delta_0,\delta_1')\|=1; \; \delta_0+\delta'_1 \Delta_1(\cdot \,| x) \geq 0 \right\}.
 \] 
Then the objective direction $ (1,-\Delta_0'(y_0\,|\,x))$ lies in the \emph{strict interior} of the dual cone $\mathbb{D}(x)^{*}$\footnote{For a set $C\subseteq\mathbb R^{L}$, its dual cone is defined as
$C^*=\{v:\langle v,d\rangle\ge 0, \forall\, d\in\operatorname{cone}(C)\}$.}: For some $\varepsilon_0>0$, 
\[
\inf_{(\delta_0,\delta_1')\in \mathbb{D}(x)}\bigl\{\, \delta_0-\delta_1'\Delta_0(y_0\,|\,x)\,\bigr\} \geq \varepsilon_0.
\]
\end{myas}

\noindent\textit{Remarks.}
Assumption~\ref{as:D} strengthens the usual ``no improving direction'' (which only requires the infimum to be $\ge 0$) by imposing a strict margin $\varepsilon_0>0$. Geometrically, $(1,-\Delta_0'(y_0\,|\,x))$ strictly separates from the recession cone: along any feasible recession direction, the objective increases at a uniformly positive rate. Together with Slater's condition (introduced later), this rules out unbounded rays in the objective direction and ensures  finiteness and robustness of the optimal value, and  a perturbation-stable, bounded solution set. The larger the margin $\varepsilon_0$, the more robust the problem is to small perturbations.

With Assumptions~\ref{as:C} and \ref{as:D}, we now examine the well-posedness of the SILP problem and the stability of \( F^{UB}_{Y_0 | D, X}(y_0 | 1, x) \).  As a first step, we verify that Slater's condition holds. This follows from the fact that, for any value \( \gamma_0 > 1 \), the vector \( (\gamma_0, 0, \dots, 0) \in \mathbb{R}^L \) lies in \( \mathbb{S}(x) \) as an interior point.  By the Strong Duality Theorem, we obtain the following result. Its proof is standard in the  SILP literature \citep[e.g.][]{goberna1998linear,hettich1993} and is therefore omitted.

\begin{lemma}\label{lem:dual}
Let \( \Lambda \) denote the set of probability measures on \( [\underline{y}, \overline{y}] \). Then strong duality holds for the semi-infinite linear program defined in Corollary~\ref{corollary1}; that is,
\begin{align*}
F^{UB}_{Y_0 | D, X}(y_0 | 1, x) = 
\max_{\lambda \in \Lambda} \quad &\int_{\underline{y}}^{\overline{y}} F_{Y | D, X}(y | 1, x) \, d\lambda(y) \\
\text{s.t.} \quad 
&\int_{\underline{y}}^{\overline{y}} \Delta_1(y | x) \, d\lambda(y) = -\Delta_0(y_0 | x).
\end{align*}
Moreover, suppose the conditions of Theorem~\ref{theorem2},  Assumption~\ref{as:C} and \ref{as:D} hold. Then both the upper bound \( F^{UB}_{Y_0 | D, X}(y_0 | 1, x) \) and the optimal solution \( \gamma^* \) are finite.
\end{lemma}

\noindent
In Lemma~\ref{lem:dual}, strong duality (under Slater) justifies the min-max Lagrangian representation (i.e., a saddle point) of the upper bound used below and guarantees dual attainment for the ensuing sensitivity analysis. It should also be noted that the optimal solution $\gamma^*$ may not be unique. Throughout, we denote  the set of optimal solutions by $\Gamma^*(x,y_0)$. Similarly, let $\lambda^*$ and $\Lambda^*(x,y_0)$ be an optimal solution to the dual problem and the collection of dual solutions, respectively.  As is standard in the LP literature, one can use the interior-point algorithm to solve the above large-scale LP dual problem (i.e. high-dimensional $\lambda$).

In practice, when the solution set is unbounded,  ``small'' but random perturbations in the LP's parameters might significantly affect  the optimal objective value. To deal with such a sensitivity issue, we introduce an additional  regularity condition.  Let $\tau>0$ and consider the following modified LP problem, denoted as $LP(\tau)$: 
\begin{align*}
  \min_{(\gamma_0,\gamma_1')\in\mathbb S(x,\tau)} & \quad \gamma_0-\gamma_1'\Delta_{0}(y_0| x),
\end{align*}
where  $\mathbb S(x,\tau)$ is defined as follows:
\[
\mathbb S(x,\tau)=\big\{\gamma\in \mathbb S(x): \|\gamma\|^2_2\leq \tau \big\}.
\]
By definition, $\mathbb S(x,\tau)$ is  a convex and compact subset of $\mathbb R^L$. 

Note that the original LP can be written as $LP(+\infty)$. By Lemma~\ref{lem:dual}, there exists a finite optimal solution $\gamma^\ast$. Hence there is a threshold $\bar\tau<\infty$ such that, for all $\tau\ge \bar\tau$, the optimal value of $LP(\tau)$ remains the same as that of the original problem. The reason is that for some optimal solution of the unregularized LP the added regularity constraint is slack, so that solution remains feasible (and optimal) in $LP(\tau)$ and the objective value is unchanged.

\begin{lemma}
\label{lemma3.1}
Consider the $LP(\tau)$ problem under the conditions of Lemma~\ref{lem:dual}. Then there exists a finite constant $\bar{\tau}_x > 0$ such that for any $\tau \geq \bar{\tau}_x$, the optimization admits at least one solution in the interior of the constraint set \( \mathbb{S}(x, \tau) \), and 
\[
F^{UB,\tau}_{Y_0 | D, X}(y_0 | 1, x) = F^{UB}_{Y_0 | D, X}(y_0 | 1, x).
\]
\end{lemma}

\noindent Given  our preceding discussion, the proof of Lemma~\ref{lemma3.1} is straightforward and therefore omitted. Throughout our following discussion, we maintain the assumption that $\tau \geq \bar{\tau}_x$.



\subsection{Nonparametric Estimation}
\label{NE}

The basic idea behind our estimation procedure is straightforward. In the \( LP(\tau) \) problem, the coefficients \( \Delta_{0z_\ell}(y_0 | x) \), \( \Delta_{1z_\ell}(\cdot | x) \), and \( F_{Y|DX}(\cdot | 1, x) \) are unknown but can be estimated nonparametrically from the data. This motivates the following two-step plug-in estimator. In the first step, we construct nonparametric estimates of these coefficients. In the second step, we solve the plug-in LP problem, denoted \( \widehat{LP}(\tau) \), under its dual formulation to obtain the estimated upper bound \( \hat{F}^{UB,\tau}_{Y_0|DX}(y_0 |1, x) \).

To clarify the idea, we consider an i.i.d. random sample \( \{(Y_i, D_i, X_i, Z_i) : i \leq n\} \) of size $n$, drawn from the population distribution of \( (Y, D, X, Z) \). 
For each \( (y, d, z, x) \in \mathcal{Y} \times \{0,1\} \times \mathcal{Z} \times \mathcal{X} \), we estimate the conditional distributions \( F_{Y|DX}(y | d, x) \) and \( F_{Y|DZX}(y | d, z, x) \) using kernel-smoothed empirical CDFs:
\begin{align*}
&\hat{F}_{Y|DX}(y | d, x) = \frac{\sum_{i=1}^n \mathbbm{1}(Y_i \leq y, D_i = d)\, K\left( \frac{x - X_i}{h_d} \right)}{\sum_{i=1}^n \mathbbm{1}(D_i = d)\, K\left( \frac{x - X_i}{h_d} \right)},\\
&\hat{F}_{Y|DZX}(y | d, z, x) = \frac{\sum_{i=1}^n \mathbbm{1}(Y_i \leq y, D_i = d, Z_i = z)\, K\left( \frac{x - X_i}{h'_d} \right)}{\sum_{i=1}^n \mathbbm{1}(D_i = d, Z_i = z)\, K\left( \frac{x - X_i}{h'_d} \right)},
\end{align*}
where \( K \) is a kernel function with compact support on $\mathbb R^{d_X}$, and \( h_0, h_1, h'_0, h'_1 \) are bandwidths associated with each conditioning group. For simplicity, we assume \( x \in \mathcal{X} \) is an interior point to avoid boundary issues in the above kernel estimates. However,  note that the boundaries on \( \mathcal{Y} \) do not affect the consistency of the kernel estimates.

Furthermore, for each \( (y, d, z, x) \in \mathcal{Y} \times \{0,1\} \times \mathcal{Z} \times \mathcal{X} \),  we estimate ${\Delta}_{dz}(y | x) $ by 
\begin{multline*}
\hat{\Delta}_{dz}(y | x) = \hat{F}_{Y|DZX}(y | d, z, x) \times \hat{p}^d(z, x)\,[1 - \hat{p}(z, x)]^{1 - d} \\
- \hat{F}_{Y|DZX}(y | d, z_L, x) \times \hat{p}^d(z_L, x)\,[1 - \hat{p}(z_L, x)]^{1 - d},
\end{multline*}
in which $\hat p(z,x)$ is a kernel estimator of $p(z,x)$, i.e., 
\[
\hat{p}(z, x) = \frac{\sum_{i=1}^n \mathbbm{1}(D_i = 1, Z_i = z)\, K\left( \frac{x - X_i}{h^\dagger} \right)}{\sum_{i=1}^n \mathbbm{1}(Z_i = z)\, K\left( \frac{x - X_i}{h^\dagger} \right)}.
\]where $h^\dag$ is also a bandwidth. 
Finally, let
\[
\hat{\Delta}_d(y | x) = \big(\hat{\Delta}_{dz_1}(y | x), \ldots, \hat{\Delta}_{dz_{L-1}}(y | x) \big).
\]

It is worth noting that the proposed estimators \( \hat{F}_{Y|DX}(\cdot | d, x) \) and \( \hat{\Delta}_{d}(\cdot | x) \) are not smooth over the support \( [\underline{y}, \overline{y}] \). In fact, they are step functions, as is typical for empirical distribution functions. As a result, \( \widehat{LP}(\tau) \) reduces to a finite-dimensional linear program, whose dimensionality increases with the sample size \( n \). Alternatively, one can construct smoothed estimators of the conditional distributions \( F_{Y|DX}(y | d, x) \) and \( F_{Y|DZX}(y | d, z, x) \) as follows:
\begin{align*}
&\tilde{F}_{Y|DX}(y | d, x) = \frac{\sum_{i=1}^n \Phi\left(\frac{y - Y_i}{b_n}\right) \mathbbm{1}(D_i = d)\, K\left( \frac{x - X_i}{h_d} \right)}{\sum_{i=1}^n \mathbbm{1}(D_i = d)\, K\left( \frac{x - X_i}{h_d} \right)}, \\
&\tilde{F}_{Y|DZX}(y | d, z, x) = \frac{\sum_{i=1}^n \Phi\left(\frac{y - Y_i}{b_n}\right) \mathbbm{1}(D_i = d, Z_i = z)\, K\left( \frac{x - X_i}{h'_d} \right)}{\sum_{i=1}^n \mathbbm{1}(D_i = d, Z_i = z)\, K\left( \frac{x - X_i}{h'_d} \right)}.
\end{align*}where \( \Phi \) denotes the standard normal CDF and \( b_n > 0 \) with \( b_n \to 0 \) as \( n \to \infty \).  With this modification, the estimated functions \( \tilde{F}_{Y|DX}(\cdot | d, x) \) and \( \tilde{F}_{Y|DZX}(\cdot | d, z, x) \) become differentiable. Interestingly, introducing such smoothness into the LP coefficients has little impact on the derivation of the asymptotic properties for   $\hat{F}^{UB,\tau}_{Y_0|DX}(y_0|1,x)$, as will be discussed below.

Furthermore, let \( \hat{\gamma}^* \) denote an optimal solution to \( \widehat{LP}(\tau) \), and let \( \hat{F}^{UB,\tau}_{Y_0|DX}(y_0 | 1, x) \) be the corresponding optimal value, i.e., 
\begin{align*}
&\hat F^{UB}_{Y_0 | D X}(y_0 | 1, x) = \min_{ \gamma \in \hat{\mathbb{S}}(x,\tau)} \, - \gamma' \hat \Delta_0(y_0 | x);\\
&\hat \gamma= \underset{ \gamma \in \hat{\mathbb{S}}(x,\tau)}{\arg\min} \, - \gamma' \hat \Delta_0(y_0 | x).
\end{align*}
Again, the optimal solution \( \hat{\gamma}^* \) may not be unique. In this case, let $\hat \Gamma^*(x,y_0)$ be the set of optimal solutions to the empirical LP problem \( \widehat{LP}(\tau) \). 

We now introduce some conditions to the above nonparametric estimators.  For notation simplicity, let $\xi( x, y_0) \equiv \big( \Delta_0(y_0 | x),\, \Delta_1(\cdot | x),\, F_{Y | D, X}(\cdot | 1, x) \big)$ be the triple of vector and functional inputs to the SILP. For any compact subset \( A \subset \mathbb{R}^d \), define
\[
\ell^\infty(A) \equiv \left\{ g : A \to \mathbb{R} \;\text{such that}\; \sup_{a \in A} |g(a)| < \infty \right\}
\]
as the space of bounded, real-valued functions on \( A \), equipped with the supremum norm. Moreover, let $\mathbb S_{\xi}\equiv  \mathbb{R}^{L-1} \times \ell^\infty([\underline{y},\, \overline{y}])^{L-1} \times \ell^\infty([\underline{y},\, \overline{y}])$ be the support of $\xi(x,y_0)$. Let further $\hat \xi( x, y_0) \equiv \big( \hat\Delta_0(y_0 | x),\, \hat\Delta_1(\cdot | x),\, \hat F_{Y | D, X}(\cdot | 1, x) \big)\in\mathbb S_\xi$ denote the estimator of $\xi(x,y_0)$.

\begin{myas}{A1}\label{as:A1}
The LP coefficients $\xi(x,y_0)$ are consistently estimated by $\hat\xi(x,y_0)$, i.e., $\hat \Delta_0(y_0 | x) - \Delta_0(y_0 | x)\overset{p}{\rightarrow} 0$ and 
\begin{align*}
&\sup_{y\in[\underline y,\overline y]}\big\| \hat \Delta_1(y | x) - \Delta_1(y | x) \big\|
\overset{p}{\rightarrow} 0;\\
&\sup_{y\in[\underline y,\overline y]} \left| \hat F_{Y| D,X}(y | 1,x) - F_{Y| D,X}(y |1,x) \right|
\overset{p}{\rightarrow} 0.
\end{align*}
\end{myas}

 While Assumption~\ref{as:A1} is  high level, it can be derived from standard primitive assumptions on, e.g., kernel functions and bandwidths in the nonparametric estimation literature; see, e.g., \citet{paganullah1999}.

\begin{myas}{A2}\label{as:A2}
For some \( \kappa \in (0, 1/2] \), we have 
\[
n^\kappa \big[
\hat \xi( x,y_0) - \xi( x,y_0)
\big] 
\rightsquigarrow  \mathbb{G}(\cdot | x,y_0)\equiv \big( \mathbb{G}_0( x,y_0), \mathbb{G}_1(\cdot | x), \mathbb{G}_F(\cdot | x) \big),
\]
where  \( \mathbb{G}_0(x,y_0 ) \in \mathbb{R}^{L-1} \) a  multivariate normal random vector, and \( \mathbb{G}_1(\cdot | x) \), \( \mathbb{G}_F(\cdot | x) \) tight Gaussian processes in \( \ell^\infty([\underline{y}, \overline{y}])^{L-1} \) and \( \ell^\infty([\underline{y}, \overline{y}]) \), respectively.
\end{myas}

Assumption~\ref{as:A2} imposes a functional central limit theorem on the first-stage estimators. 
Since weak convergence in $\ell^\infty$ entails stochastic equicontinuity, we obtain, for any $\epsilon_n \downarrow 0$, 
\begin{align*} 
\sup_{\|y'-y\|\leq \epsilon_n} \left\|\hat{\mathbb G}_1(y'|x) - \hat{\mathbb G}_1(y|x)\right\| &= o_p(1), \\
\sup_{\|y'-y\|\leq \epsilon_n} \left|\hat{\mathbb G}_F(y'|x) - \hat{\mathbb G}_F(y|x)\right| &= o_p(1),
\end{align*}
where $\hat{\mathbb G}_{1}(\cdot|x) = n^{\kappa}\big[\hat \Delta_1(\cdot | x) - \Delta_1(\cdot | x)\big]$ and 
$\hat{\mathbb G}_F(\cdot|x) = n^{\kappa}\big[\hat F_{Y| D,X}(\cdot | 1,x) - F_{Y| D,X}(\cdot |1,x)\big]$. 
These properties are standard in the empirical process literature and can be established under primitive conditions 
such as smoothness of the underlying distributions and regularity of kernel or series estimators; 
see, e.g., \citet{vandervaart1996weak} and \citet{hardle2004nonparametric}.



\subsection{Numerical Studies\label{sec:Numerical-Studies}}
\label{subsec:mc-multiIV}

To illustrate (i) how having multiple instrument values sharpens our bounds and (ii) our two-step SILP estimation procedure, we conduct a focused Monte Carlo exercise. We consider the following DGP that satisfies Conditions~\ref{as:condi_1} but violates \ref{as:condi_0}. We generate a random sample  $\{(Y_i,D_i,Z_i): i\leq n\}$ of $(Y,D,Z)$ generated as follows:
\begin{align*}
 &Y = 1-D + (1+D)U_D, \\
& D = \mathbbm{1}(\eta\, \leq\, \pi_0+\pi_1 Z ),
\end{align*}with $(\pi_0,\pi_1)=(0.2,0.5)$. For $d=0,1$, let further 
\[
U_d = U + \xi_d,\quad \ \xi_0=\xi_1+\nu,
\] with (i) $(\nu,\,\xi_1) \bot\,  (U,\eta)$; (ii) $\nu\bot\, \xi_1$;  (iii) both $\nu$ and $\xi_1$ conform to standard normal distributions; (iv)  $(U,\eta)$ conform to joint normal with mean $(0,0)'$ and variance $\Sigma=[1,0.8; 0.8,1]$.
 Moreover, let $Z\bot \, (\nu,  \xi_1,U, \eta)$ and $Z \sim \text{Bin}(L-1,p)/(L-1) \;\in [0,1]$ for $L\in \{2,3,4\}$ and $p=0.5$. Here, $Z$ is normalized so that the endpoints of the support are invariant regardless of the
value of $L$. This is intended to understand the role of the number of values Z takes while
fixing the role of instrument strength fixed. 

\begin{figure}[!htbp]
\centering
\includegraphics[width=0.4\linewidth]{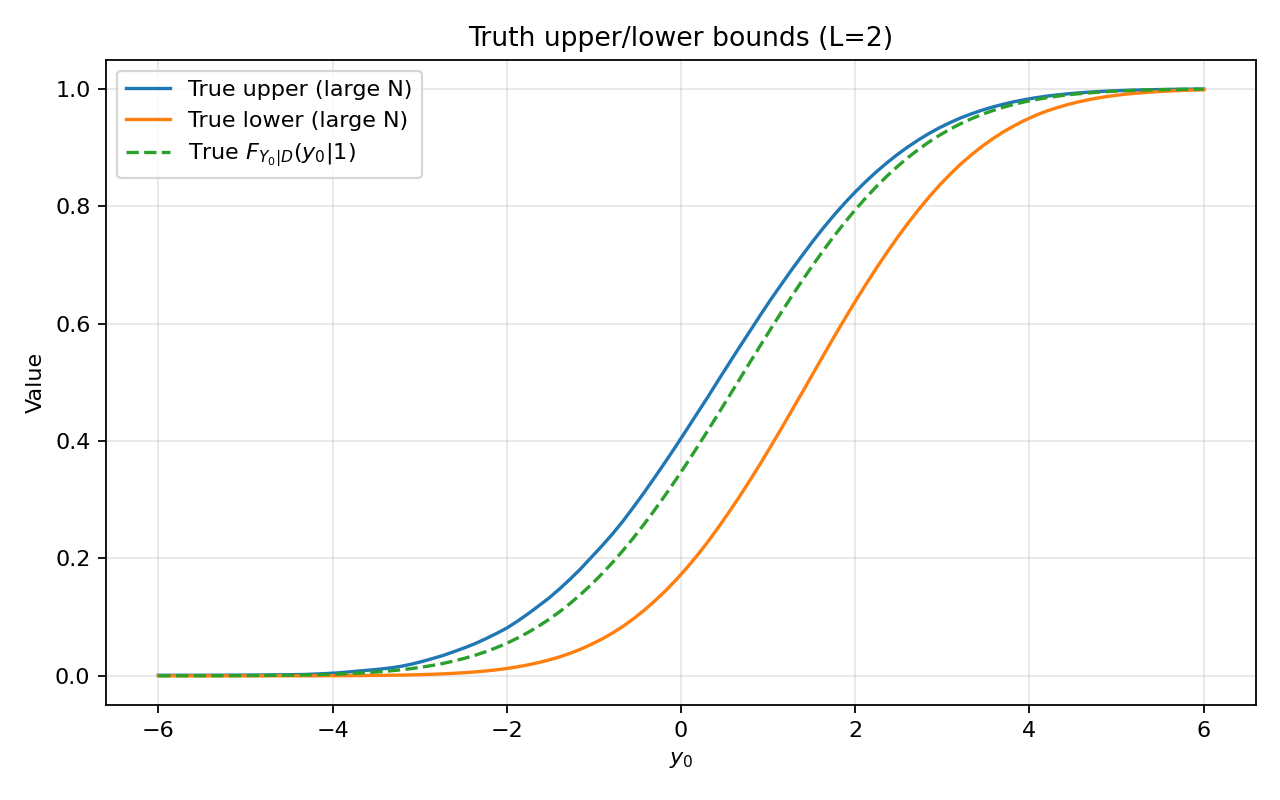}
\includegraphics[width=0.4\linewidth]{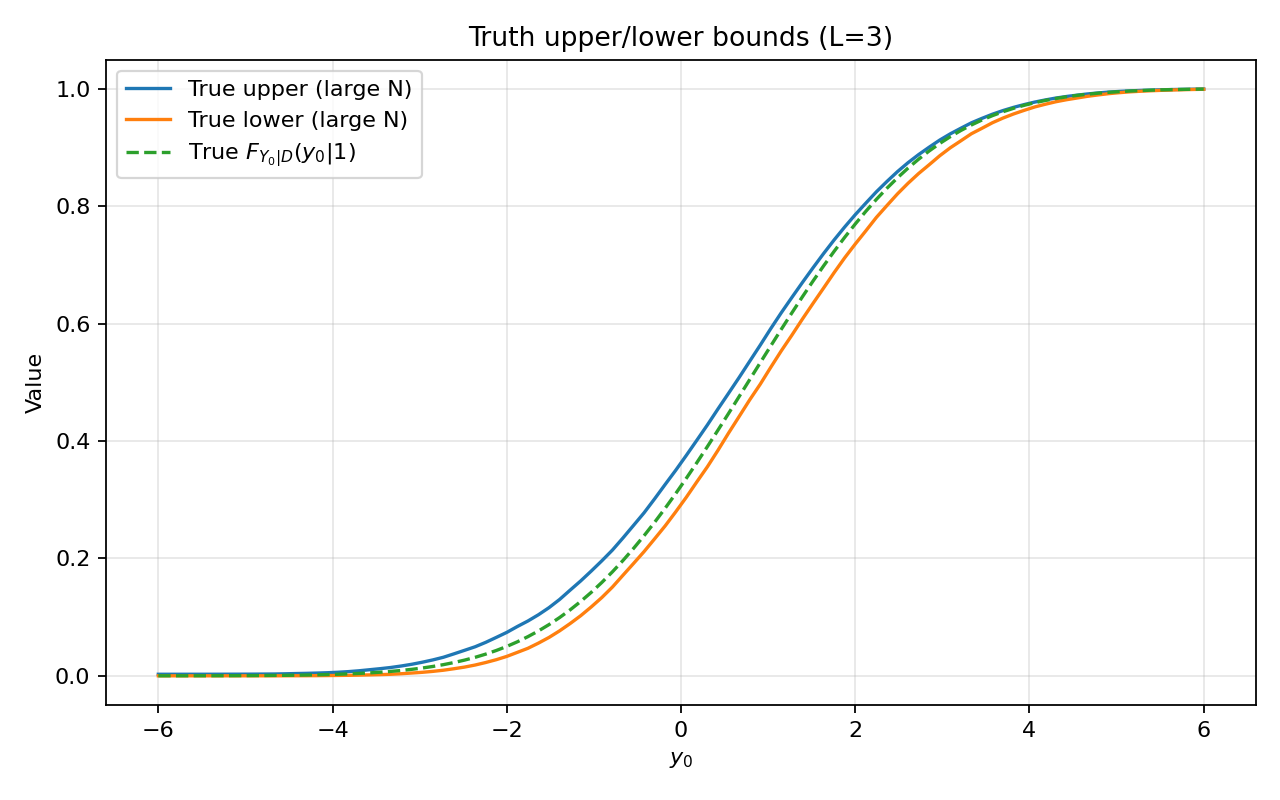}\\
\includegraphics[width=0.4\linewidth]{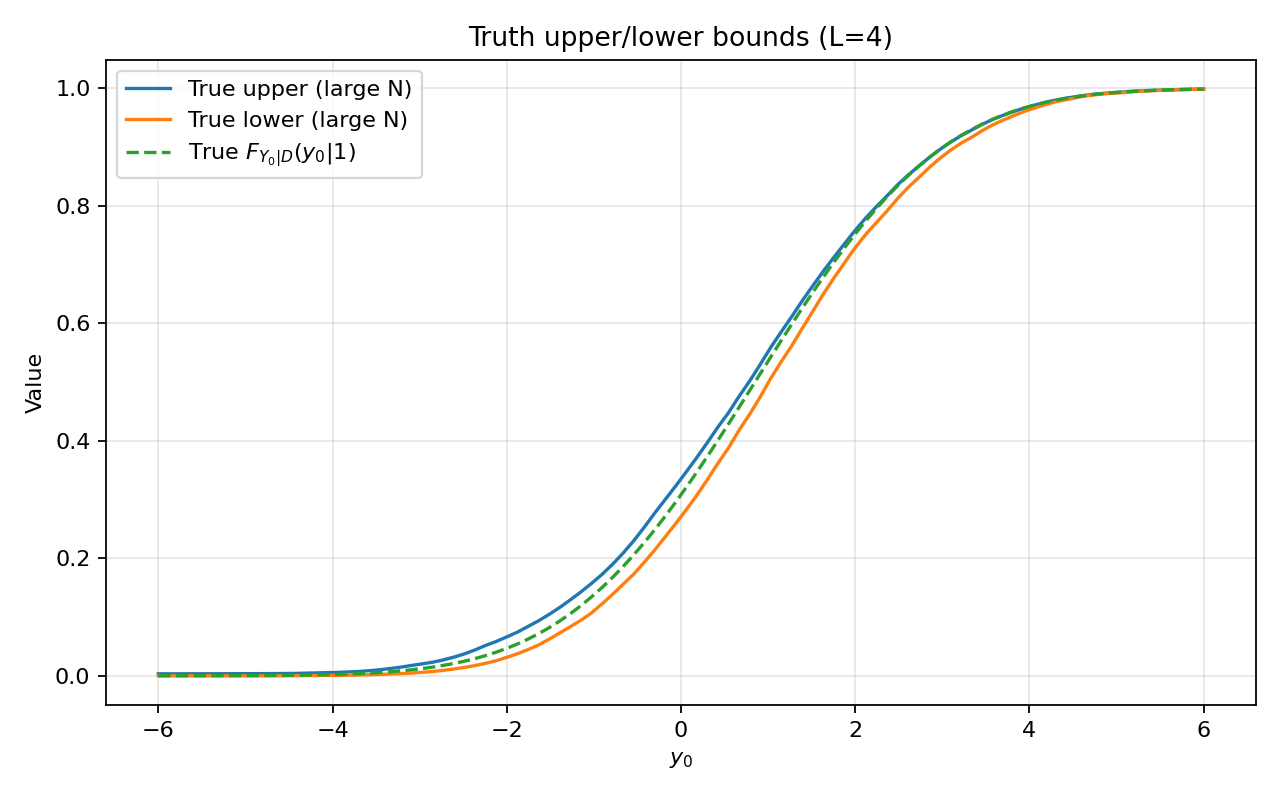}
\includegraphics[width=0.4\linewidth]{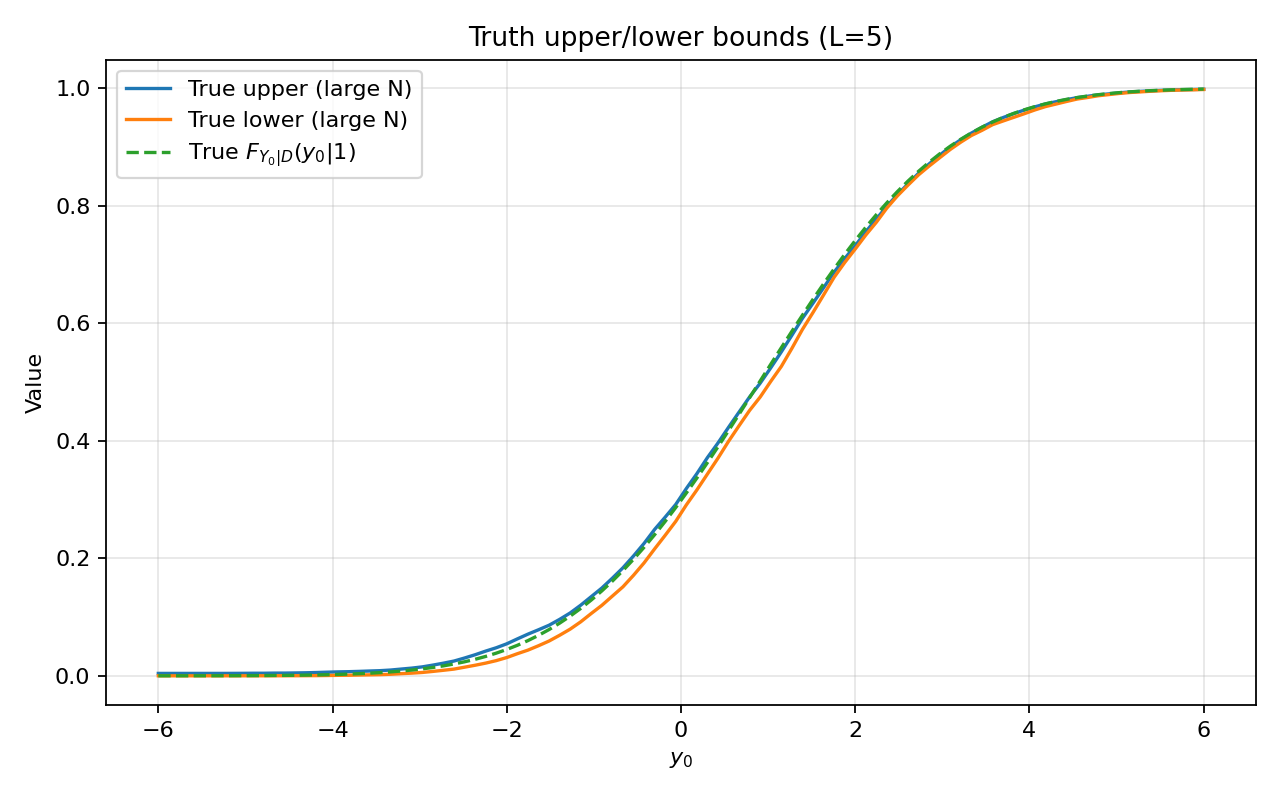}
\caption{Bounds on $\,\Pr(Y_{0}\le \cdot| D=1)$ with $L\in\{2,3,4,5\}$}
\label{fig1}
\end{figure}

Our object of interest is the distribution of the potential untreated outcome among the treated, $F_{Y_0 | D}(\cdot \;| 1)$, and we compute identified upper and lower bounds as defined in Corollary~\ref{corollary1}. 
Figure~\ref{fig1} displays the true $F_{Y_0 | D}(\cdot \;| 1)$ together with the estimated lower and upper bounds for $L \in \{2,3,4,5\}$.  All curves are obtained from simulations with a large sample size ($n=10{,}000{,}000$). 
The tuning parameter is fixed at $\tau=100$, but the regularization constraint $\|\gamma\| \le \tau$ is slack at the optimum and thus does not affect the SILP solutions (with a few exceptions due to sampling error). 
The upper and lower bounds are computed pointwise for each $y_0 \in [-6,6]$. 
As $L$ increases, i.e., as the IV offers richer support, the bounds tighten.

Fix $L=2$. We now examine the finite sample performance of the proposed SILP estimator. Specifically, we set $N=1000,2000$ and $4000$. For each $N$, we run $R=200$ replications to estimate the mean of the upper/lower bound  and then construct 95\% CI. To ensure Assumption~\ref{as:D} hold for a sufficient large $\varepsilon_0$, we focus on $y_0\in[-0.520,\,1.688]$, where the two endpoints correspond to the 25th and 75th quantiles of $F_{Y_0| D}(\cdot\,| 1)$, respectively. For $y_0\notin[-0.520,\,1.688]$, note that the objective direction is close to the boundary of the dual cone of  the feasible region's recession directions. Accordingly, rather than solving the SILP problems, we take the set of optimal solutions computed on $y_0\in[-0.520,\,1.688]$ and optimize the objective function over that set rather than over the SILP's feasible region. The resulting minimum and maximum objective values are reported as the upper and lower bounds outside the highlighted interval $[-0.520,\,1.688]$, respectively.

\begin{figure}[!htbp]
\centering
\includegraphics[width=0.4\linewidth]{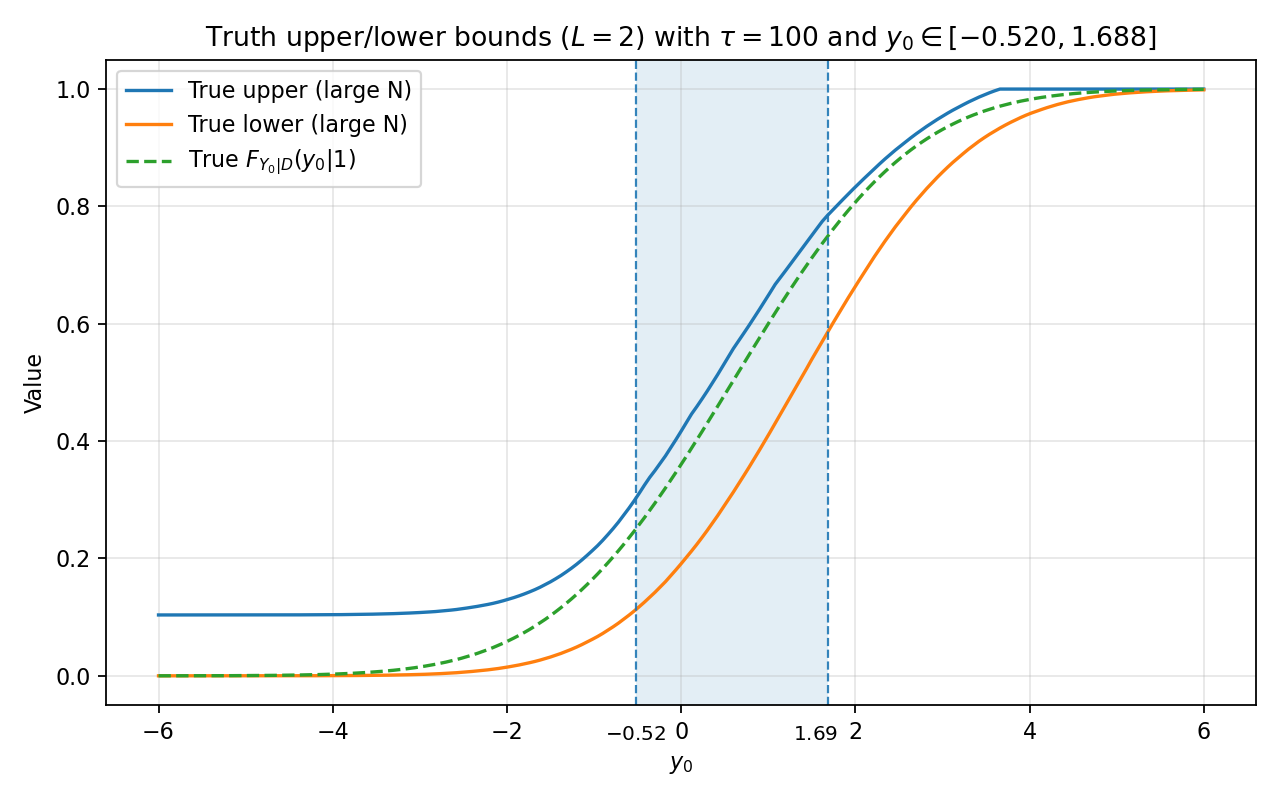}
\includegraphics[width=0.4\linewidth]{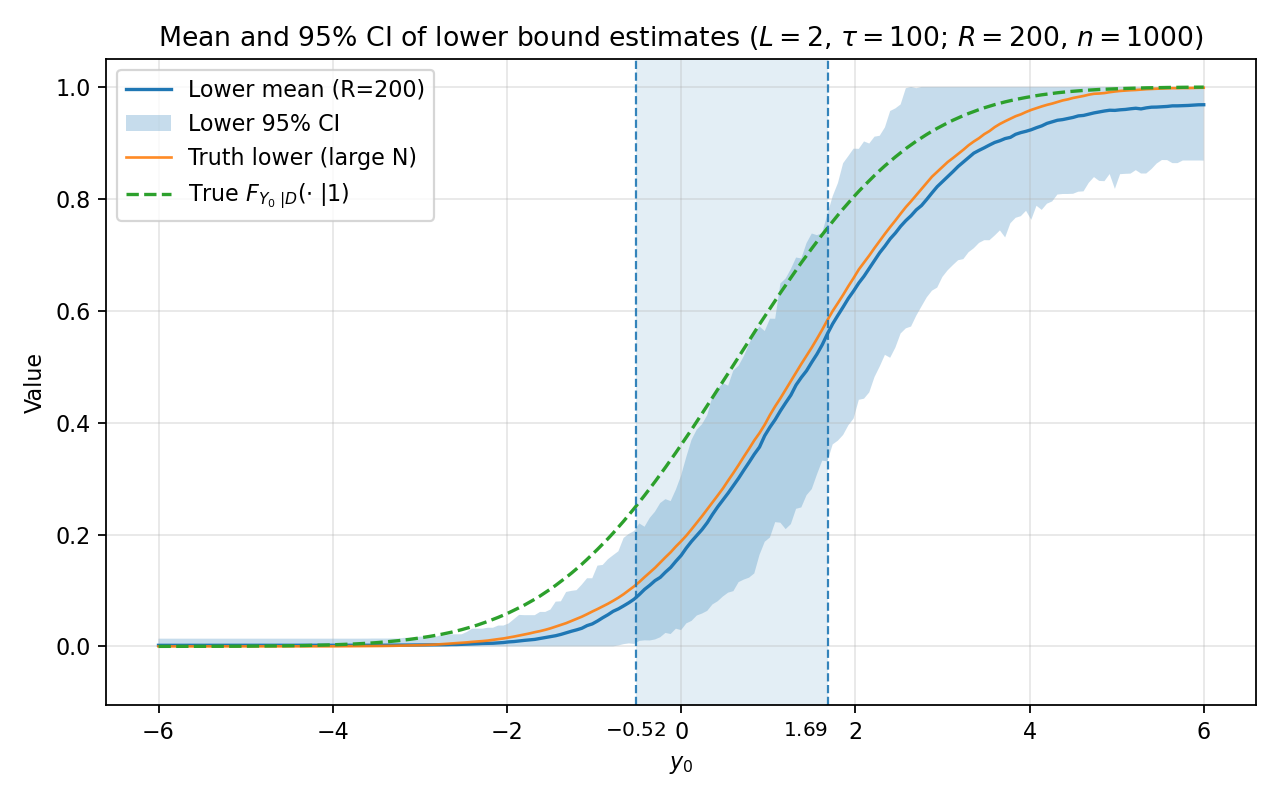}\\
\includegraphics[width=0.4\linewidth]{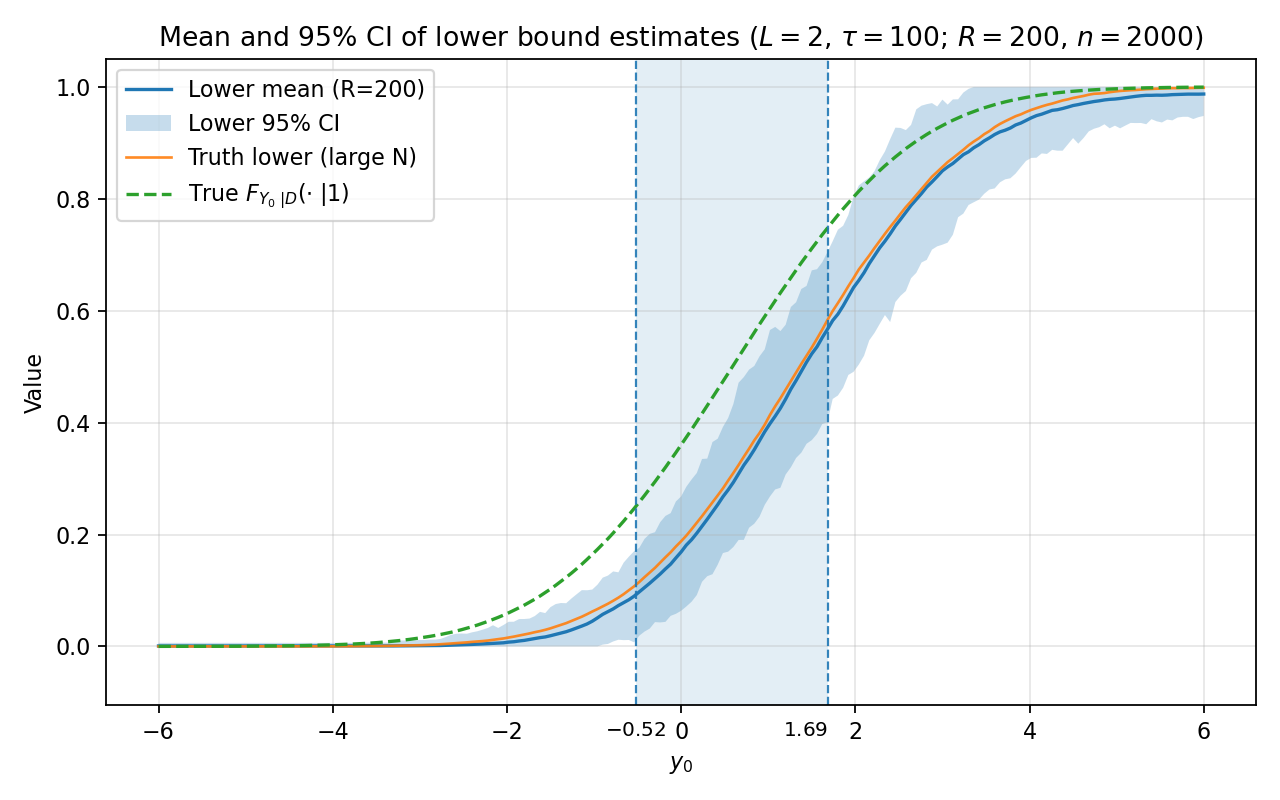}
\includegraphics[width=0.4\linewidth]{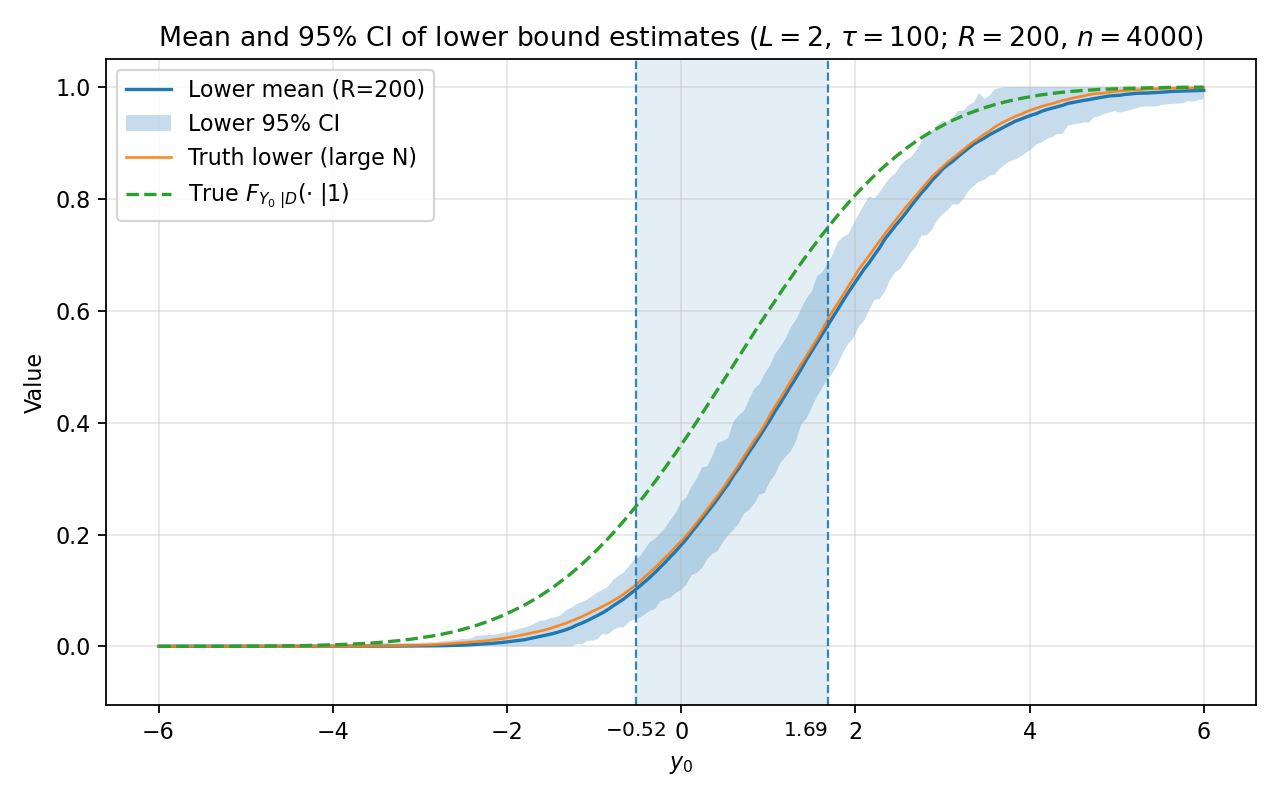}
\caption{Lower bound estimates with mean and  95\% CI}
\label{fig2}
\end{figure}
\begin{figure}[!htbp]
\centering
\includegraphics[width=0.4\linewidth]{fig_truth_anchor_L2.png}
\includegraphics[width=0.4\linewidth]{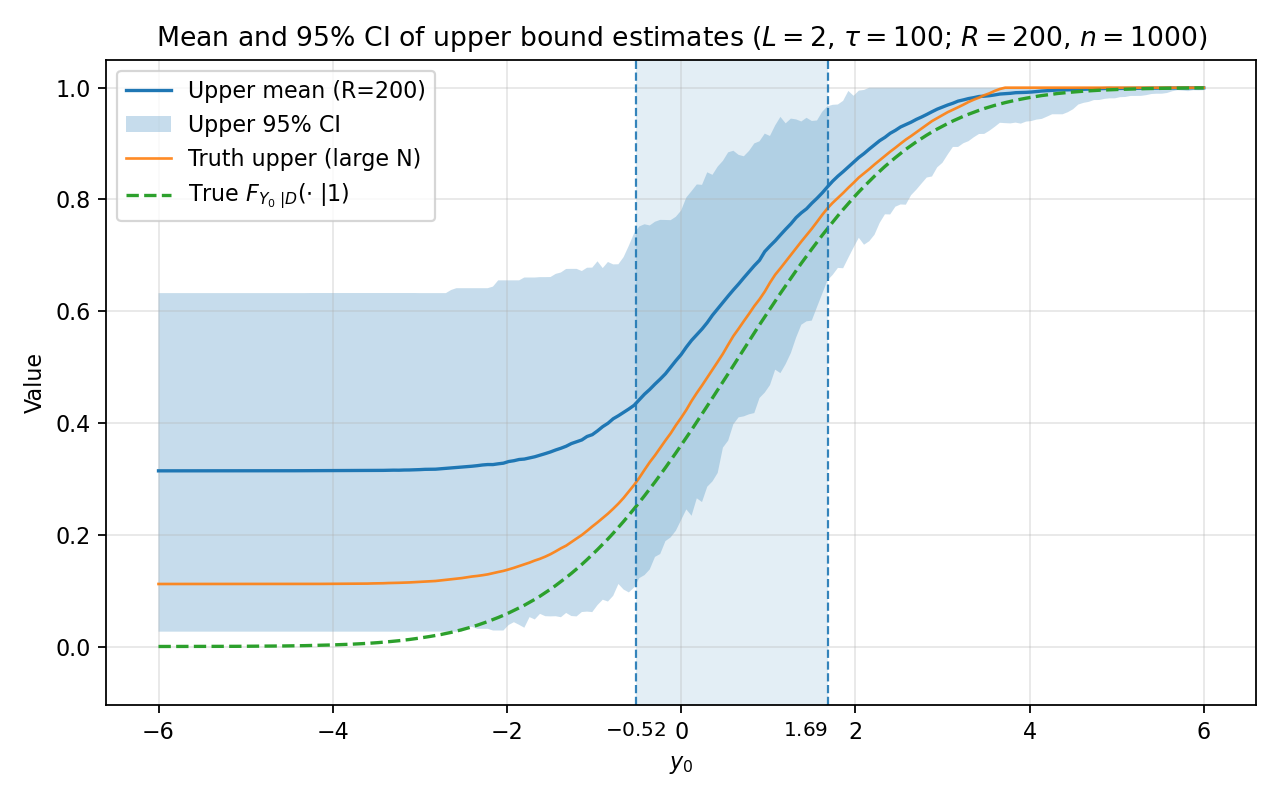}\\
\includegraphics[width=0.4\linewidth]{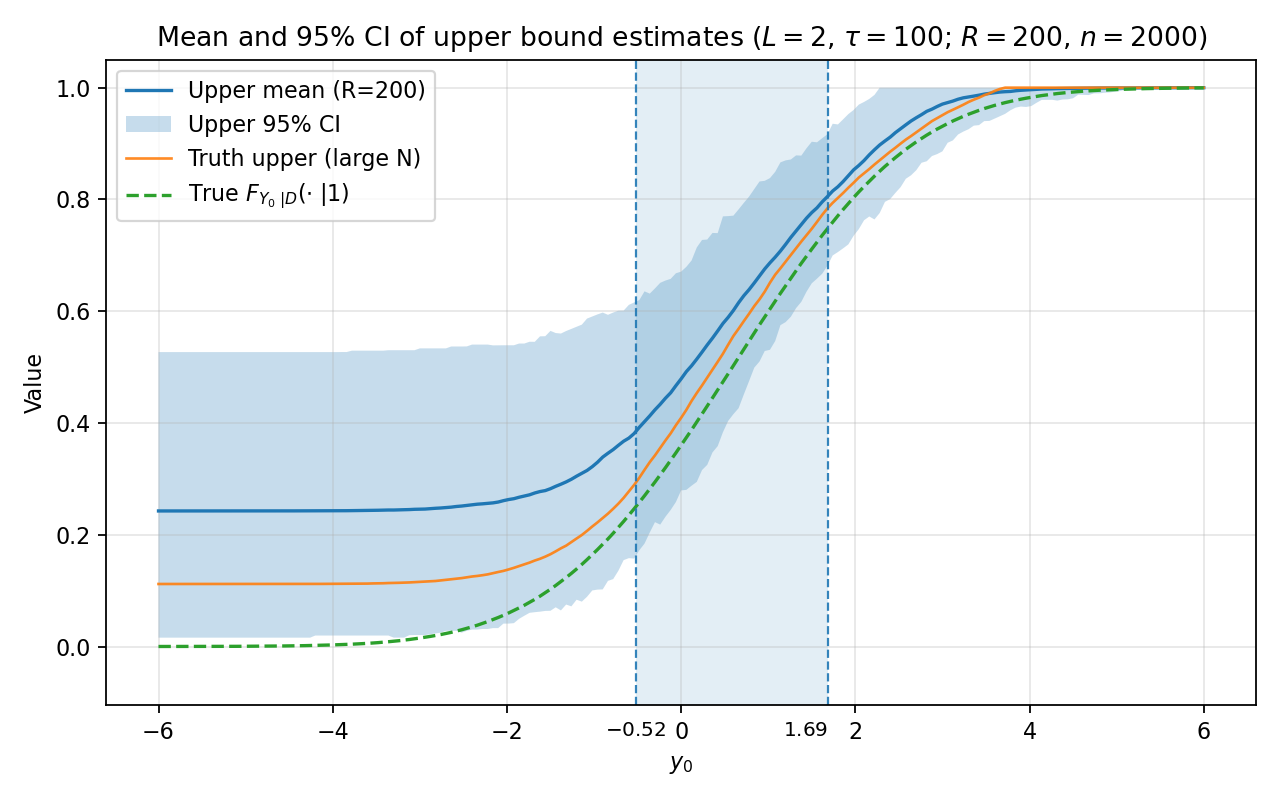}
\includegraphics[width=0.4\linewidth]{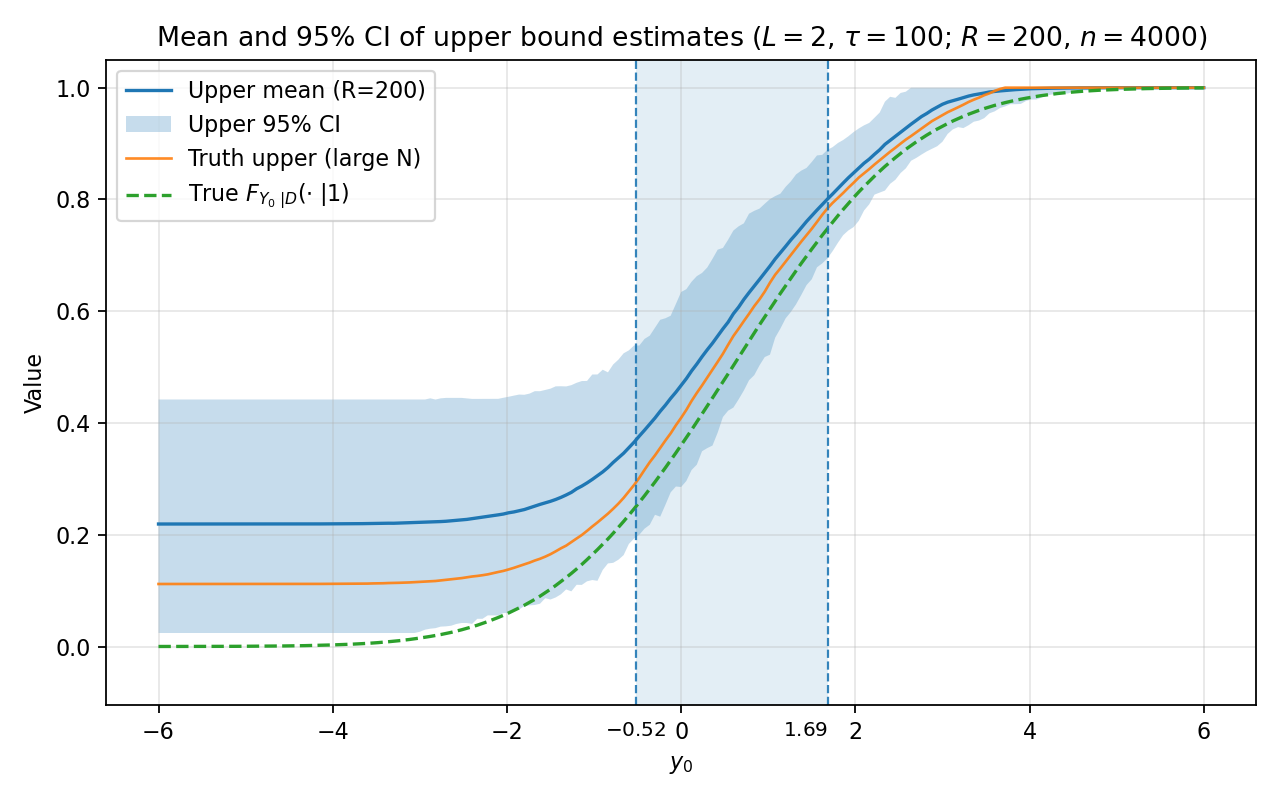}
\caption{Lower bound estimates with mean and  95\% CI}
\label{fig3}
\end{figure}

Figures~\ref{fig2} and~\ref{fig3} display the means and pointwise 95\% CIs of the SILP bounds based on $R=200$ replications. Within the highlighted interval $[-0.520,\,1.688]$, we find: (i) the true bound lies within the confidence band; (ii) the mean of the lower bound estimator perfectly matches the true lower bound, while the mean of the upper bound estimator converges to the true upper bound as the sample size increases\footnote{For the upper bound problem, Assumption~\ref{as:D} holds in a weak sense for small values of $y_0$.}; and (iii) all the dispersions of CIs decrease with sample size. 

\section{Asymptotic Properties\label{sec asy}}
Section~\ref{sec:Systematic-Calculation-of} introduces a nonparametric estimator  \( \hat{F}^{UB,\tau}_{Y_0 \mid D, X}(\cdot | 1, x) \) for the upper bound \( {F}^{UB,\tau}_{Y_0 \mid D, X}(\cdot | 1, x) \) by solving an empirical  SILP. In this section, we study the asymptotic properties of this estimator. In addition, we also consider the large sample behavior of the associated empirical optimizer $\hat\gamma^*$ when the solution is unique.  Throughout, our analysis is \emph{pointwise} in the covariate
value, i.e., fixing \( X = x \). Extending the theory to process convergence over
the support of \(X\)  is left for future work.

\subsection{Consistency and Weak Convergence}

The asymptotic analysis of empirical SILP estimators has been studied by establishing Hadamard directional differentiability of the value map with respect to small perturbations of the SILP coefficients; see, e.g., \citet{shapiro1991asymptotic} and \citet{bonnans2000perturbation}. Under Slater's condition and compactness (in the weak-* topology) of the primal and dual optimal sets, $\Gamma^*(x,y_0)\subseteq\mathbb{R}^L$ and $\Lambda^*(x,y_0)\subseteq\Lambda$, they show that, for convex programs (including SILPs), the value function is Hadamard directionally differentiable, and that its derivative admits a min-max representation over the associated optimal sets.\footnote{Note that $\Lambda=\mathcal P([\underline y,\overline y])$ is defined as the set of Borel probability measures on the compact interval $[\underline y,\overline y]$, endowed with the topology of weak convergence (i.e., $\lambda_n\Rightarrow\lambda$ iff $\int f\,d\lambda_n\to\int f\,d\lambda$ for all $f\in\mathcal C([\underline y,\overline y])$). Because $[\underline y,\overline y]$ is a compact metric space, $\Lambda$ is compact in this topology (by Prokhorov's theorem). Note that the dual problem imposes linear moment constraints, which are weakly closed, hence the dual-feasible set is a closed subset of $\Lambda$ and therefore compact.} Following their approach, we treat our upper-bound estimator as a min-max object and begin by establishing its consistency.


For fixed $(x,y_0)$, define the Lagrangian
\[
\mathcal{L}\big(\xi(x,y_0);\gamma,\lambda\big)
\equiv  \gamma_0 - \gamma_1' \Delta_0(y_0| x)
+ \int_{\underline y}^{\overline y}\!\big[F_{Y| D,X}(y| 1,x) - \gamma_0 - \gamma_1' \Delta_1(y| x)\big]\,d\lambda(y),
\]
with $\gamma\in\mathcal B_\tau\equiv\{\gamma\in\mathbb R^L:\|\gamma\|\le\tau\}$ and $\lambda\in\Lambda$.  
Define the value functional $\phi:\mathbb S_\xi\to\mathbb R$ by
\[
\phi(\xi)\;\equiv\;\min_{\gamma\in\mathcal B_\tau}\ \max_{\lambda\in\Lambda}\ \mathcal L(\xi;\gamma,\lambda).
\]
By Lemma~\ref{lem:dual} (strong duality) and Sion's minimax theorem, since $\mathcal B_\tau$ is compact, $\Lambda$ is weakly compact \citep[see e.g.][]{billingsley1999convergence}, and $\mathcal L$ is jointly continuous, convex in $\gamma$, and concave  in $\lambda$,  \cite{sion1958general}'s minimax theorem gives
\[
F^{UB,\tau}_{Y_0| D,X}(y_0| 1,x)
=\min_{\gamma\in\mathcal B_\tau}\max_{\lambda\in\Lambda}\mathcal L\big(\xi(x,y_0);\gamma,\lambda\big)
=\max_{\lambda\in\Lambda}\min_{\gamma\in\mathcal B_\tau}\mathcal L\big(\xi(x,y_0);\gamma,\lambda\big).
\]
See, e.g., \cite{rockafellar1970convex}. It implies that any $(\gamma^*,\lambda^*)\in\Gamma^*\times\Lambda^*$ is a saddle point satisfying
\[
\mathcal L(\xi(x,y_0);\gamma^*,\lambda)\ \le\ 
\mathcal L(\xi(x,y_0);\gamma^*,\lambda^*)\ \le\ 
\mathcal L(\xi(x,y_0);\gamma,\lambda^*)\quad
\forall(\gamma,\lambda)\in\mathcal B_\tau\times\Lambda,
\]
which we use to establish continuity of $\phi$ (hence consistency) and, under additional regularity, its Hadamard directional derivative.

\begin{theorem}\label{theorem:4.1} 
Suppose the conditions of Lemma~\ref{lem:dual} and Assumption~\ref{as:A1} hold. Then
\[
\hat{F}^{UB,\tau}_{Y_0| D,X}(y_0| 1,x)\ \overset{p}{\rightarrow}\ F^{UB,\tau}_{Y_0| D,X}(y_0| 1,x).
\]
\proof See Appendix~\ref{proof_theorem4.1}.\qed
\end{theorem}

We next derive the limiting distribution of $\hat{F}^{UB,\tau}_{Y_0|D,X}(y_0\,|\,1,x)$. 
By the generalized envelope theorem of \citet{milgrom2002envelope},
the value functional $\phi$ is Hadamard directionally differentiable at $\xi(x,y_0)$. The derivative takes the form of a support-function-type expression over the primal-dual optimal sets $(\Gamma^*,\Lambda^*)$ and is therefore, in general, only piecewise linear in the perturbation direction.

\begin{lemma}\label{lem:4.2}
Suppose the conditions of Lemma~\ref{lem:dual} hold. Define
\[
\phi(\xi)= \min_{\gamma\in\mathcal B_\tau}\ \max_{\lambda\in\Lambda}\ \mathcal L(\xi;\gamma,\lambda),
\quad \xi\in\mathbb S_{\xi}.
\]
Then $\phi$ is Hadamard directionally differentiable at $\xi=\xi(x,y_0)$: for any direction $\delta=(\delta_0,\delta_1,\delta_F)\in\mathbb S_{\xi}$,
\begin{equation}\label{eq:silp-derivative}
\phi'_H(\xi(x,y_0);\delta)=
\min_{\gamma\in\Gamma^*}\ \max_{\lambda\in\Lambda^*(\gamma)}
\left\{-\,\gamma_1' \delta_0\;+\;\int_{\underline y}^{\overline y}\!\big[\delta_F(y)-\gamma_1'\delta_1(y)\big]\,d\lambda(y)\right\},
\end{equation}
where
\[
\Lambda^*(\gamma)=\arg\max_{\lambda\in\Lambda}\ \mathcal L(\xi(x,y_0);\gamma,\lambda).
\]
\proof see Appendix~\ref{proof_lemma4.1}. \qed
\end{lemma}



Under Assumption~\ref{as:A2}, note that the empirical process $n^{\kappa}[\hat\xi(x,y_0)-\xi(x,y_0)]$ converges in distribution to $\mathbb G_\xi(\cdot|x,y_0)=(\mathbb G_0(x,y_0),\mathbb G_1(\cdot|x),\mathbb G_F(\cdot|x))$,  a tight Gaussian element in the space $\mathbb S_{\xi}$. For each $(\gamma,\lambda)\in\mathcal B_\tau\times\Lambda$, define 
\[
\mathbb Z(\gamma,\lambda|x,y_0)\equiv 
-\gamma'_1\mathbb G_0(x,y_0)
\;+\;\int_{\underline y}^{\overline y}\!\Big[\mathbb G_F(y| x)-\gamma_1' \mathbb G_1(y| x)\Big]\,d\lambda(y).
\]
Because $(\gamma,\lambda)\mapsto \mathbb Z(\gamma,\lambda| x,y_0)$ is a continuous linear functional of $\mathbb G_\xi(\cdot|x,y_0)$ and the index set $\mathcal B_\tau\times\Lambda$ is compact, the process $\{\mathbb Z(\gamma,\lambda| x,y_0):(\gamma,\lambda)\in\mathcal B_\tau\times\Lambda\}$ is a tight Gaussian process.

\begin{theorem}\label{theorem:asymptotic-distribution}
Fix $X=x$ and $\tau\ge\bar\tau_x$. Suppose the conditions of Lemma~\ref{lem:dual} and Assumptions~\ref{as:A1}-\ref{as:A2} hold. Then
\[
n^{\kappa} \Big[ \hat{F}^{UB,\tau}_{Y_0 | D,X}(y_0 | 1,x) - F^{UB,\tau}_{Y_0 |D,X}(y_0 | 1,x) \Big]
\overset{d}{\rightarrow}
\min_{\gamma\in\Gamma^*}\ \max_{\lambda\in\Lambda^*}\ \mathbb Z(\gamma,\lambda\,|\,x,y_0).
\]
\end{theorem}
\noindent
If the optimal set $\Gamma^*$ or $\Lambda^*$ is not a singleton, the limiting distribution is a min-max functional of a Gaussian process, reflecting the piecewise-linear geometry of the value functional $\phi$. In particular, $\phi$ is only \emph{Hadamard directionally} differentiable at $\xi(x,y_0)$, its derivative map is generally nonlinear, rather than fully (linearly) Hadamard differentiable. As emphasized by \citet{fang2019inference}, this lack of full differentiability can render the standard bootstrap invalid. In such cases one typically employs nonstandard methods for inference (e.g., the derivative-based bootstrap in \citealp{fang2019inference} and the numerical delta method in \citealp{hong2018numerical}).


%

\subsection{Local Stability and Asymptotic Normality}
In this subsection, we establish asymptotic normality of the SILP's optimal value and, under uniqueness and nondegeneracy, of the optimal solution. Specifically, we impose additional regularity ensuring \emph{local stability} of the primal and dual solutions, so that the solution mapping is locally unique and Lipschitz in $\xi(x,y_0)$. Under these additional conditions, the value functional $\phi$ is \emph{fully Hadamard differentiable} at $\xi(x,y_0)$ (i.e., it admits a continuous linear derivative), in contrast to the merely directional differentiability that holds without such stability.

In the LP sensitivity literature, \emph{local stability} means that under small perturbations of the coefficients $\xi(x,y_0)$, the set of binding constraints and their multipliers remain stable, and the optimal primal and dual solutions vary smoothly (e.g., locally Lipschitz). A standard sufficient condition is the Linear Independence Constraint Qualification (LICQ) \citep[see][]{bonnans2000perturbation}, which, in our SILP with optimaization vector $\gamma\in\mathbb R^L$, requires exactly $L$ linearly independent active constraints at the optimum $\gamma^*$. This can be overly restrictive and lacks clear economic motivation in our application. We therefore impose a weaker regularity condition that still guarantees local stability while allowing the number of active constraints to  be equal or less than $L$.

\begin{myas}{R}\label{as:R}
There exists a unique solution \(\gamma^* \in \mathbb{R}^{L}\) to \(LP(\tau)\).
Moreover, at \(\gamma^*\) there are exactly \(K\) binding constraints (with \(1 \le K \le L\)) at distinct points \(\{y_{1k}^*\}_{k=1}^K \subset [\underline y,\overline y]\), and the corresponding dual multipliers satisfy \(\lambda_k^*>0\) for all \(k=1,\ldots,K\).
Assume further that the \(K\) binding constraints are linearly independent.
\end{myas}

In Assumption~\ref{as:R}, linear independence of the active constraints, together with strictly positive dual multipliers (a form of strict complementarity), ensures uniqueness of the dual solution. To see this, note that the first-order KKT conditions gives us\footnote{Note that the constraint $\|\gamma\|^2\leq \tau$ is inactive.}
\[
\Delta_{0}(y_0| x)+\sum_{k=1}^K \lambda_k^*\,\Delta_{1}(y^*_{1k}| x)=0.
\]
Because the \(K\) active constraints (i.e. linear equations with slopes \(\{[\,1;\,\Delta_1(y^*_{1k}| x)\,]\}_{k=1}^K\)) are linearly independent, the above linear system admits a unique solution for the strictly positive dual variables \(\{\lambda_k^*\}_{k=1}^K\). W.l.o.g., we assume  that the first \( K \) components of the coefficient vectors in the  active constraints are linearly independent, i.e., 
\[
Rank \begin{pmatrix}
1 & \Delta_{11}(y_{1,1}^* | x) & \cdots & \Delta_{1,K-1}(y_{1,1}^* | x) \\
\vdots & \vdots &  & \vdots \\
1 & \Delta_{11}(y_{1,K}^* | x) & \cdots & \Delta_{1,K-1}(y_{1,K}^* | x)
\end{pmatrix}=K.
\]

Under Assumption~\ref{as:R}, the asymptotic normality of the estimated upper bound follows directly from Theorem~\ref{theorem:asymptotic-distribution}. Let \((\gamma^*,\lambda^*)\) be the unique optimal primal-dual solution pair. Then the limiting distribution in Theorem~\ref{theorem:asymptotic-distribution} is normal, i.e.,
\[
n^{\kappa}\Big[\hat{F}^{UB,\tau}_{Y_0| D,X}(y_0| 1,x)-F^{UB,\tau}_{Y_0| D,X}(y_0| 1,x)\Big]
\ \overset{d}{\rightarrow}\ \mathbb{Z}(\gamma^*,\lambda^*\,|x,y_0).
\]
This shows that the way first-stage estimation error propagates into sampling variation of the bound estimator is governed by the optimal primal vector \(\gamma_1^*\) and the active-constraint multipliers \(\{\lambda_k^*\}_{k=1}^K\).

Next, we  move to the asymptotic properties for the  estimator $\hat\gamma^*$ as a solution to the empirical SILP problem. Recall that the optimal solution $\gamma^*$ identifies the mixture of \emph{complier} subgroups: their potential treated outcome $Y_1$ FOSD  the probability distribution of treated outcome. 

With a unique primal-dual solution under Assumption~\ref{as:R}, we establish consistency of \(\hat\gamma^*\) by adapting the Argmin Theorem \citep[see e.g.][]{vandervaart1998asymptotic} to the saddle-point (min-max) structure:
For each \(\gamma\in\mathcal B_\tau\), denote
\begin{align*}
\lambda^\dag(\gamma)\in\arg\max_{\lambda\in\Lambda}\ \mathcal L\big(\xi(x,y_0);\gamma,\lambda\big),\quad \text{and } \;
\hat\lambda^\dag(\gamma)\in\arg\max_{\lambda\in\Lambda}\ \mathcal L\big(\hat\xi(x,y_0);\gamma,\lambda\big).
\end{align*}
For notation simplicity, let $\xi=\xi(x,y_0)$ and $\hat \xi=\hat \xi(x,y_0)$ in the following discussion. By Assumption~\ref{as:A1}, we obtain
\begin{align*}
0
&\le \mathcal L\big(\xi;\hat\gamma,\lambda^\dag(\hat\gamma)\big)-\mathcal L\big(\xi;\gamma^*,\lambda^\dag(\gamma^*)\big)\\
&=\big[\mathcal L\big(\hat\xi;\hat\gamma,\lambda^\dag(\hat\gamma)\big)-\mathcal L\big(\xi;\gamma^*,\lambda^\dag(\gamma^*)\big)\big]
+ \big[\mathcal L\big(\xi;\hat\gamma,\lambda^\dag(\hat\gamma)\big)-\mathcal L\big(\hat\xi;\hat\gamma,\lambda^\dag(\hat\gamma)\big)\big] \\
&\leq \mathcal L\big(\hat\xi;\hat\gamma,\hat\lambda^\dag(\hat\gamma)\big)-\mathcal L\big(\xi;\gamma^*,\hat\lambda^\dag(\gamma^*)\big) + o_p(1) \\
&=\big[\mathcal L\big(\hat\xi;\hat\gamma,\hat\lambda^\dag(\hat\gamma)\big)-\mathcal L\big(\hat\xi;\gamma^*,\hat\lambda^\dag(\gamma^*)\big)\big]
 + \big[\mathcal L\big(\hat\xi;\gamma^*,\hat\lambda^\dag(\gamma^*)\big)-\mathcal L\big(\xi;\gamma^*,\hat\lambda^\dag(\gamma^*)\big)\big]
 + o_p(1) \\
&\le o_p(1),
\end{align*}
where all the three inequalities use the saddle point structure. 
By uniqueness of \(\gamma^*\) and compactness of \(\mathcal B_\tau\), the standard Argmin Theorem gives \(\hat\gamma\overset{p}{\to}\gamma^*\). By a similar argument, we can also show that $\hat\lambda^*\overset{p}{\rightarrow}\lambda^*$ in the space $\Lambda$ (which is compact in the weak-* topology). 

Furthermore, we consider the limiting distribution of $\hat\gamma^*$. Because there are \(K\) active constraints (\(K\le L\)) at the optimum, this suggests that there might be less binding constraints at $\hat\gamma^*$ than the SILP's dimensionality $L$. To deal with this issue, we restructure the SILP problem via a two-step nested optimization. 
First, we reparametrize the optimization vector \(\gamma\in\mathbb R^L\) as \(\gamma=(\theta_1,\theta_2)\in\mathbb R^K\times\mathbb R^{L-K}\), where 
\(
\theta_1=(\gamma_0,\gamma_{11},\dots,\gamma_{1,K-1})'\in\mathbb R^K\) and 
\(\theta_2=(\gamma_{1K},\dots,\gamma_{1,L-1})'\in\mathbb R^{L-K}\).  For each $y\in[\underline y, \overline y]$ and $d\in\{0,1\}$, let further $\Psi_{d1}(y| x)=\big[(-1)^{d+1},\ \Delta_{d1}(y| x),\dots,\Delta_{d,K-1}(y| x)\big]' \in\mathbb R^{K}$, and $\Psi_{d2}(y| x)=\big[\Delta_{dK}(y| x),\dots,\Delta_{d,L-1}(y| x)\big]' \in\mathbb R^{L-K}$. Moroever, for each fixed \(\theta_2\), we define an inner SILP problem, denoted as $\widetilde {LP}(\theta_2,\tau)$, for solving the \(K\)-dimensional parameter \(\theta_1\) as follows:
\[
\begin{aligned}
Q(\theta_2| x,y_0)
=\ &\underset{\theta_1\in {\mathcal B}_{1\tau}}{\min}\quad  -\theta_{1}'\Psi_{01}(y_0| x)- \theta_2'\,\Psi_{02}(y_0| x)\\
\text{s.t.}\quad 
& F_{Y| D,X}(y| 1,x) \le \theta_{1}'\Psi_{11}(y| x)+ \theta_2'\Psi_{12}(y| x),\quad \forall\, y\in[\underline y,\overline y],
\end{aligned}
\]
where \({\mathcal B}_{1\tau}=\{\theta_1\in\mathbb R^{K}:\|\theta_1\|^2\leq \tau\}\). In addition, let \(\theta_1^\dagger(\theta_2)\) be the solution to the inner problem. Under Assumption~\ref{as:R}, $\theta_1^\dagger(\theta_2)$ is well defined for $\theta_2$ belongs to a neighborhood of $\theta_2^*$, i.e., there exists a unique solution to the above inner SILP problem,  when $\theta_2\in\mathcal N_\epsilon(\theta^*_2)$ for some $\epsilon>0$. By definition, $Q(\cdot|x,y_0)$ is convex.\footnote{To see this, let  $\rho\in[0,1]$, and $\theta_2,\tilde \theta_2\in \tilde B_\tau$. Note that $\theta_{1,\rho}\equiv \rho\theta_1^\dag(\theta_2)+(1-\rho)\theta^\dag(\tilde \theta_2)$  belongs to the feasible region of  $\widetilde {LP}(\rho\theta_2+(1-\rho)\tilde \theta_2,\tau)$, and 
\[
\rho Q(\theta_2|x,y_0)+(1-\rho) Q(\tilde \theta_2|x,y_0)=  \theta'_{1,\rho}\Psi_{01}(y_0|x)\geq Q(\rho\theta_2+(1-\rho)\tilde \theta_2|x,y_0).
\]}

In the outer step, we minimize the criterion \(Q(\cdot| x,y_0)\) over the compact set \( B_{2\tau}=\{\theta_2\in\mathbb R^{L-K}:\ \|\theta_2\|^2\le \tau\}\), i.e.
\[
\min_{\theta_2\in  B_{2\tau}}\ Q(\theta_2| x,y_0).
\]
By Assumption~\ref{as:R}, the unique solution \(\gamma^*=(\theta_1^*,\theta_2^*)\) to the population SILP is recovered by the two-step procedure: \(\theta_2^*\) solves the outer problem and \(\theta_1^*=\theta_1^\dag(\theta_2^*)\) solves the inner SILP. Appendix~B establishes local stability of the inner SILP at its optimum, covering \(\theta_1^\dag(\theta_2)\), the active set, and the associated multipliers for \(\theta_2\) in a neighborhood \(\mathcal N_\varepsilon(\theta_2^*)\). In addition, we also provide an expression for the first and second order derivative of $Q(\cdot|x,y_0)$ by applying \citet{milgrom2002envelope}'s generalized envelope theorem and exploiting the KKT conditions.

Under local stability (unique optimizer, constant active set, smooth multipliers), the generalized envelope theorem of \citet{milgrom2002envelope} gives a Hadamard directional derivative for the inner value, which in our setting  is linear in \(\theta_2\) within a neighborhood of \(\theta_2^*\). Therefore, \(Q(\cdot| x,y_0)\) is continuously differentiable at \(\theta_2^*\). In particular, since \(\theta_2^*\) uniquely minimizes \(Q(\cdot| x,y_0)\), we have $\frac{\partial Q(\theta_2^*| x,y_0)}{\partial \theta_2}=0$, and the Hessian matrix \(\frac{\partial^2 Q(\theta_2^*| x,y_0)}{\partial \theta_2\partial \theta_2'}\) is positive semi-definite. In Appendix~B, we derive the expressions for $\frac{\partial \hat Q(\theta_2|x,y_0)}{\partial \theta_2}$ and $\frac{\partial Q(\theta_2^*| x,y_0)}{\partial \theta_2\partial \theta_2'}$  by applying \citet{milgrom2002envelope}'s generalized envelope theorem.


\begin{lemma}\label{lem:4.6}
Fix $X=x$ and $\tau\ge\bar\tau_x$. Suppose the conditions of Lemma~\ref{lem:dual}, Assumptions~\ref{as:A1}, \ref{as:A2}, and \ref{as:R} hold. Suppose in addition  that the Hessian matrix \(\frac{\partial^2 Q(\theta_2^*| x,y_0)}{\partial \theta_2\partial \theta_2'}\) has full rank. Then
\[
n^{\kappa}(\hat\theta_2-\theta^*_2)=-\left[\frac{\partial^2  Q(\theta_2^* | x,y_0)}{\partial \theta_2 \partial \theta_2'}\right]^{-1}\times n^{\kappa}  \frac{\partial \hat Q(\theta^*_2|x,y_0)}{\partial \theta_2}+o_p(1),
\]where  \(n^{\kappa}\frac{\partial\hat Q(\theta_2^*\mid x,y_0)}{\partial \theta_2}\) is asymptotically normal.

\end{lemma}
\noindent
Using Taylor expansion of the outer FOC gives us the stated representation for \(n^{\kappa}(\hat\theta_2-\theta_2^*)\). The proof is straightforward and therefore omitted.

Next, we derive the asymptotic distribution of $\hat\theta_1^*$ by exploiting the KKT conditions from the inner SILP. In particular, we  characterize how the estimated binding points $\{\hat y^*_{1k}\}$ and the constraint residuals (slacks) respond to perturbations in the estimated coefficients $\hat\xi(x,y_0)$.

For $(\theta_1,\theta_2)\in\mathbb R^{K}\times\mathbb R^{L-K}$ and $y_1\in[\underline y,\overline y]$, define the \emph{residual functions}, which is associated with the constraints,  as follows: 
\[
\begin{aligned}
r(y_1 | x, \theta_1,\theta_2) 
&= F_{Y | D,X}(y_1 | 1, x) - \theta_{1}' \Psi_{11}(y_1 | x) - \theta_2' \Psi_{12}(y_1 | x), \\
\hat{r}(y_1 | x, \theta_1,\theta_2) 
&= \hat{F}_{Y | D,X}(y_1 | 1, x) - \theta_{1}' \hat{\Psi}_{11}(y_1 | x) - \theta_2' \hat\Psi_{12}(y_1 | x).
\end{aligned}
\]
Note that $\hat r$ is the plug-in estimator of $r$. At the primal optima, feasibility implies  $r(\cdot\,|x,\theta_1^*,\theta^*_2)\leq 0$ and $\hat r(\cdot\,|x,\hat\theta_1^*,\hat\theta^*_2)\leq 0$; at binding points, we have $r(y_1^*\,|\,x,\theta_1^*,\theta^*_2)=0$ and $\hat r(\hat y_1^*\,|x,\hat\theta_1^*,\hat \theta^*_2)=0$.

By definition, it follows immediately that
\begin{align*}
\hat r(\hat y^*_{1k} |x, \hat \theta_1,\hat \theta_2) 
 - r(\hat y^*_{1k} | x, \theta^*_1, \theta^*_2) &\geq 0, \\ 
\hat r(y^*_{1k} | x, \hat \theta_1,\hat \theta_2) 
 - r(y^*_{1k} | x, \theta^*_1, \theta^*_2) &\leq 0. 
\end{align*}
These inequalities highlight the distance between the estimated and population residual functions at the 
active constraints. Moreover,  because for any $y_1 \in [\underline y, \overline y]$, 
\[
\hat r(y_1 | x, \hat \theta_1,\hat \theta_2) 
= \hat r(y_1 | x, \theta^*_1, \theta^*_2)  - (\hat\theta_{1}-\theta^*_{1})' \hat{\Psi}_{11}(y_1 | x) - (\hat\theta_2-\theta^*_2)' \hat \Psi_{12}(y_1 | x),
\]
therefore, substituting this expression into the inequalities above gives us: 
\begin{align*}
\hat r(\hat y^*_{1k} | x, \theta^*_1,\theta^*_2) 
 - r(\hat y^*_{1k} | x, \theta^*_1, \theta^*_2) 
&\geq (\hat\theta_{1}-\theta^*_{1})' \hat{\Psi}_{11}(\hat y^*_{1k} | x) 
   + (\hat\theta_2-\theta^*_2)' \hat \Psi_{12}(\hat y^*_{1k} | x), \\ 
\hat r(y^*_{1k} | x,  \theta^*_1, \theta^*_2) 
 - r(y^*_{1k} |x, \theta^*_1, \theta^*_2) 
&\leq (\hat\theta_{1}-\theta^*_{1})' \hat{\Psi}_{11}(y^*_{1k} | x) 
   + (\hat\theta_2-\theta^*_2)' \hat \Psi_{12}(y^*_{1k} | x). 
\end{align*}
By  the equicontinuity assumed  in 
Assumption~\ref{as:A2}, we obtain the following result.

\begin{lemma}
\label{lem:4.5}
Fix $X=x$ and $\tau\ge\bar\tau_x$. Suppose all the conditions in Lemmas~\ref{lem:dual},  and Assumptions~\ref{as:A1}, \ref{as:A2} and  \ref{as:R} hold.  Then, for each $k \leq K$, we have
\[
\hat{r}\big( y^*_{1k} | x, \theta^*_1, \theta^*_2 \big)
 - {r}\big( y^*_{1k} | x, \theta^*_1, \theta^*_2 \big)
= (\hat \theta_{1}-\theta^*_{1})' \Psi_{11}(y^*_{1k} | x)
  + (\hat\theta_2-\theta^*_2)' \Psi_{12}( y^*_{1k} | x)
  + o_p(n^{-\kappa}).
\]
\end{lemma}
\noindent
The proof is omitted, as it follows directly from the previous discussion.

For notational convenience, let  $y_1^* \equiv (y_{11}^*,\ldots,y_{1K}^*)'\in [\underline y, \, \overline y]^K$ and define 
\[
{r}\big( y^*_{1} | x, \theta^*_1, \theta^*_2 \big)=[{r}\big( y^*_{11} | x, \theta^*_1, \theta^*_2 \big)
;\cdots; {r}\big( y^*_{1K} | x, \theta^*_1, \theta^*_2 \big)]
\]as a $K$-dimensional column vector. Let further 
\begin{align*}
&\mathbb A_{11}(y_1^*|x)
= \big[\Psi_{11}(y^*_{11}|x),\dots,\Psi_{11}(y^*_{1K}|x)\big];\\
&\mathbb A_{12}(y_1^*|x)
= \big[\Psi_{12}(y^*_{11}|x),\dots,\Psi_{12}(y^*_{1K}|x)\big].
\end{align*} By definition, $\mathbb A_{11}(y_1^*|x)\in\mathbb R^{K\times K}$ and $\mathbb A_{12}(y_1^*|x)\in\mathbb R^{(L-K)\times K}$ denote the Jacobian matrices of the active constraints w.r.t. $\theta_1$ and $\theta_2$, respectively. Note that we assume  $\mathbb A_{11}(y_1^*|x)$ has full rank, hence is invertible.

\begin{theorem}
\label{thm:theta1-asymp}
Fix $X=x$ and $\tau\ge\bar\tau_x$. Suppose the conditions of Lemma~\ref{lem:dual}, Assumptions~\ref{as:A1}, \ref{as:A2}, and \ref{as:R} hold. Suppose in addition  that the Hessian matrix \(\frac{\partial^2 Q(\theta_2^*| x,y_0)}{\partial \theta_2\partial \theta_2'}\) has full rank. Then
\begin{multline*}
n^\kappa(\hat \theta_{1}-\theta^*_{1})
=\left[ \mathbb A'_{11}(y^*_{1} | x) \right]^{-1}\times n^\kappa\left[\hat{r}\big( y^*_{k} | x, \theta^*_1, \theta^*_2 \big)
 - {r}\big( y^*_{k} | x, \theta^*_1, \theta^*_2 \big)\right]\\
+\left[ \mathbb A'_{11}(y^*_{1} | x) \right]^{-1} \mathbb A'_{12}( y^*_{1} | x)\times\left[\frac{\partial^2  Q(\theta_2^* | x,y_0)}{\partial \theta_2 \partial \theta_2'}\right]^{-1} \times n^\kappa \frac{\partial \hat Q(\theta^*_2|x,y_0)}{\partial \theta_2}
  + o_p(n^{-\kappa}),
\end{multline*}
which converges in distribution to a limiting normal distribution.
\end{theorem}
\section{Application: Return to Education}
TBA

\clearpage

\clearpage
\clearpage{}

\begin{appendix}
\small
\section{Proofs\label{sec:Proofs}}

\subsection{Proof of Lemma \ref{lemma1}}
\label{proof_lemma1}
\proof We first show the first half of Lemma \ref{lemma1}. Suppose RS holds. By the strict monotonicity of $q(d,x,\cdot)$, 
\[
\int_\mathcal T F_{Y_{d}|X, \eta}(\cdot| x,t)dG(t)=\int_\mathcal T F_{U_{d}|X, \eta}(q^{-1}(d,x,\cdot)| x,t)dG(t).
\]It follows that 
\begin{multline*}
\int_\mathcal T F_{Y_{1}| X, \eta}(q(1,x,\cdot)| x,t)dG(t)\leq c\quad \Longleftrightarrow \quad \int_\mathcal T F_{U_{1}|X, \eta}(\cdot| x,t)dG(t)\leq c\\
\overset{by RS}{\Longleftrightarrow}\quad \int_\mathcal T F_{U_{0}|X, \eta}(\cdot| x,t)dG(t)\leq c\quad \Longleftrightarrow\quad \int_\mathcal T F_{Y_{0}|X, \eta}(q(0,x,\cdot)| x,t)dG(t)\leq c.
\end{multline*}Therefore, Condition \ref{condi_2} holds. 

We now show the second half of the lemma. By the assumption,  for each $u\in[0,1]$, there exists  an absolutely continuous  function \( W_{u}: \mathcal{T} \to \mathbb{R} \) such that
\[
\int_{\mathcal T} F_{U_1|X,\eta}(\cdot| x,t) \, dW_u(t) = \int_{\mathcal T} F_{U_0|X,\eta}(\cdot|x,t) \, dW_{u}(t) =\mathbf{1}(\cdot \geq {u}).
\]
For any absolutely continuous  function $G:\mathcal T\rightarrow \mathbb R$, there exists a sequence $\{(\delta_k,\mu_k)\in\mathbb R^2: k=1,\cdots,\infty\}$ such that  $\sum_{k=1}^K\delta_{k}\mathbf 1 (\cdot\leq u_k)$ is monotone increasing in $K$ and 
\[
 \int F_{U_1|X,\eta}(\cdot|x,t) \, dG(t)=\lim_{K\rightarrow\infty}\sum_{k=1}^K\delta_{k}\mathbf 1 (\cdot\leq u_k)
 = \int_{\mathcal T} F_{U_1|X,\eta}(\cdot|x,t) \, d \left\{\lim_{K\rightarrow\infty}\sum_{k=1}^K\delta_{k} W_{u_k}(t) du\right\},
\]where the last step applies the Monotone Convergence Theorem.   By Condition \ref{condi_2}, we have 
\[
 \int F_{U_0|X,\eta}(\cdot|x,t) \, dG(t)
 = \int_{\mathcal T} F_{U_0|X,\eta}(\cdot|x,t) \, d \left\{\lim_{K\rightarrow\infty}\sum_{k=1}^K\delta_{k} W_{u_k}(t) du\right\}=\lim_{K\rightarrow\infty}\sum_{k=1}^K\delta_{k}\mathbf 1 (\cdot\leq u_k).
\]It follows that
\[
 \int F_{U_1|X,\eta}(\cdot|x,t) \, dG(t)= \int F_{U_0|X,\eta}(\cdot|x,t) \, dG(t).
\]
Let $G$ be  probability mass distributions, then we have 
\[
F_{U_1|X,\eta}(\cdot|x,t) = F_{U_0|X,\eta}(\cdot|x,t). \qed
\]
\subsection{Proof of Lemma \ref{lemma:2.2}}
\label{proof_lemma2.2}

\proof
Suppose 
\[
\int_\mathcal T F_{Y_{1}|X, \eta}(\cdot\,| x,t)dG(t)\leq c.
\] Note that this inequality implies that we should have $c\geq 0$, since the LHS equals to zero at $-\infty$.
By the strict monotonicity of $q(1,x,\cdot)$, the above condition implies that  
\[
\int_\mathcal T F_{U_{1}|X, \eta}(\cdot\,| x,t)dG(t)\leq c.
\]
Let $H(u_1|x)\equiv \int_\mathcal T  F_{U_1|X,\eta}(u_1|x,t)\,dG(t)$. By definition, $H(\cdot|x)\leq c$ and $H(0|x)= 0$. Then 
\begin{eqnarray*}
&&\int_\mathcal T F_{U_{0}|X, \eta}(\cdot|x,t)dG(t)=\int_0^1 F_{U_{0}|U_1,X}(\cdot|u_1, x) d H(u_1|x)\\
&=&F_{U_{0}|U_1X}(\cdot|1, x) H(1|x)-\int_0^1 H(u_1|x)dF_{U_{0}|U_1,X}(\cdot|u_1, x) \\
&\leq& c\times F_{U_{0}|U_1X}(\cdot|1, x) - c\times \int_0^1 dF_{U_{0}|U_1,X}(\cdot|u_1, x) \\
&=&c \times F_{U_{0}|U_1X}(\cdot|0, x)
\leq c,
\end{eqnarray*}
where the first inequality comes from  $\frac{dF_{U_{0}|U_1X}(\cdot|u_1, x)}{du_1}\leq 0$ under Assumption \ref{as:I}. 

Again by the the strict monotonicity of $q(0,x,\cdot)$, we have
\[
\int_\mathcal T F_{Y_{0}|X, \eta}(\cdot| x,t)dG(t)\leq c.
\]

We now prove the second half of the lemma. Suppose Condition (i) and (ii) hold and 
\[
\int_\mathcal T F_{Y_{0}|X, \eta}(\cdot| x,t)dG(t)\leq c.
\]
Note that for any $u_0\in[0,1]$
\begin{multline*}
0\geq \int_\mathcal T F_{U_{0}|X \eta}(u_0|x,t)dG(t)-c=\int_0^1 H(u_1|x)\times \left\{-\frac{\partial F_{U_{0}|U_1X}(u_0|u_1, x)}{\partial u_1} \right\}du_1 -c\\
= \int_0^1 [H(u_1| x)-c]\times \left\{-\frac{\partial F_{U_{0}|U_1X}(u_0| u_1, x)}{\partial u_1} \right\}du_1
\end{multline*}where both equalities use Condition (i). The above inequality implies that $H(\cdot| x)-c\leq 0$ holds almost everywhere on $[0,1]$, as  under Condition (ii), any positive measure set where 
would contradict the inequality for some $u_0\in[0,1]$.\qed

\subsection{Proof of Theorem \ref{theorem2}}
\label{proof_theorem2}
\proof
By Condition \ref{as:condi_1}, \eqref{eq2.3} implies that 
\begin{multline*}
\Pr(Y_0\leq\cdot|D=1,X=x)\leq\gamma_0+\sum_{\ell=1}^{L}\gamma_{\ell}\Pr(Y_0\leq\cdot,D=1|Z=z_{\ell},X=x)\\
=\gamma_0+\sum_{\ell=1}^{L}\gamma_{\ell}\Pr(Y_0\leq\cdot\mid Z=z_{\ell},X=x)-\sum_{\ell=1}^{L}\gamma_{\ell}\Pr(Y_0\leq\cdot; D=0\mid Z=z_{\ell},X=x)\\
=\gamma_0+\sum_{\ell=1}^{L}\gamma_{\ell}\Pr(Y_0\leq\cdot\mid X=x)-\sum_{\ell=1}^{L}\gamma_{\ell}\Pr(Y\leq\cdot; D=0\mid Z=z_{\ell},X=x)\\
=\gamma_0-\sum_{\ell=1}^{L}\gamma_{\ell}\Pr(Y\leq\cdot; D=0\mid Z=z_{\ell},X=x),
\end{multline*} where the last uses the definition of $\gamma_L$, which implies that $\sum_{\ell=1}^{L}\gamma_{\ell}=0$.
\qed

\subsection{Proof of Theorem~\ref{theorem:4.1}}
\label{proof_theorem4.1}
\proof
Note that $F^{UB,\tau}_{Y_0| D,X}(y_0| 1,x)=\phi(\xi)$ and
$\hat F^{UB,\tau}_{Y_0| D,X}(y_0| 1,x)=\phi(\hat\xi)$. By Assumption~\ref{as:A1},
\[
 \sup_{(\gamma,\lambda)\in\mathcal B_\tau\times\Lambda}
\big|\mathcal L(\hat\xi(x,y_0);\gamma,\lambda)-\mathcal L(\xi(x,y_0);\gamma,\lambda)\big|=o_p(1).
\]
Because for any $(\gamma,\lambda)\in \mathcal B_\tau\times\Lambda$,
\[
\mathcal L(\hat\xi(x,y_0);\gamma,\lambda)\le \mathcal L(\xi(x,y_0);\gamma,\lambda)+ \sup_{(\gamma,\lambda)\in\mathcal B_\tau\times\Lambda}
\big|\mathcal L(\hat\xi(x,y_0);\gamma,\lambda)-\mathcal L(\xi(x,y_0);\gamma,\lambda)\big|,
\]
it follows that 
\begin{multline*}
\phi(\hat\xi(x,y_0))\ =\ \min_{\gamma\in\mathcal B_\tau}\max_{\lambda\in\Lambda} \mathcal L(\hat\xi(x,y_0);\gamma,\lambda)\\
\leq \ \min_{\gamma\in\mathcal B_\tau}\max_{\lambda\in\Lambda} \Big\{\mathcal L(\xi(x,y_0);\gamma, \lambda)+ \sup_{(\gamma,\lambda)\in\mathcal B_\tau\times\Lambda}
\big|\mathcal L(\hat\xi(x,y_0);\gamma,\lambda)-\mathcal L(\xi(x,y_0);\gamma,\lambda)\big|\Big\}\\
=\min_{\gamma\in\mathcal B_\tau}\max_{\lambda\in\Lambda} \mathcal L\big(\xi(x,y_0);\gamma,\lambda\big)+ \sup_{(\gamma,\lambda)\in\mathcal B_\tau\times\Lambda}
\big|\mathcal L(\hat\xi(x,y_0);\gamma,\lambda)-\mathcal L(\xi(x,y_0);\gamma,\lambda)\big|.
\end{multline*}
Similarly, we have
\[
\mathcal L(\xi(x,y_0);\gamma,\lambda)\le \mathcal L(\hat\xi(x,y_0);\gamma,\lambda)+ \sup_{(\gamma,\lambda)\in\mathcal B_\tau\times\Lambda}
\big|\mathcal L(\hat\xi(x,y_0);\gamma,\lambda)-\mathcal L(\xi(x,y_0);\gamma,\lambda)\big|.
\] which implies that 
\[
\phi(\xi(x,y_0))\leq\phi(\hat\xi(x,y_0))+ \sup_{(\gamma,\lambda)\in\mathcal B_\tau\times\Lambda}
\big|\mathcal L(\hat\xi(x,y_0);\gamma,\lambda)-\mathcal L(\xi(x,y_0);\gamma,\lambda)\big|.
\]
 Hence
\[
|\phi(\hat\xi)-\phi(\xi)|\ \le\  \sup_{(\gamma,\lambda)\in\mathcal B_\tau\times\Lambda}
\big|\mathcal L(\hat\xi(x,y_0);\gamma,\lambda)-\mathcal L(\xi(x,y_0);\gamma,\lambda)\big|=o_p(1).\qed
\]

\subsection{Proof of Lemma~\ref{lem:4.2}}
\label{proof_lemma4.1}
\proof
Fix $(x,y_0)$. Consider Hadamard perturbations $\xi_\varepsilon(x,y_0)=\xi(x,y_0)+\varepsilon\delta_\varepsilon$ with $\delta_\varepsilon\to\delta$ in $\mathbb S_{\xi}$ as $\varepsilon\downarrow0$. Since $\mathcal L$ is affine in $\xi(x,y_0)$,
\[
\mathcal L(\xi_\varepsilon(x,y_0);\gamma,\lambda)
=\mathcal L(\xi(x,y_0);\gamma,\lambda)+ \varepsilon\times \bigg\{ -\gamma'_1 \delta_0\;+\;\int_{\underline y}^{\overline y}\!\big[\delta_F(y)-\gamma'_1 \delta_1(y)\big]\,d\lambda(y)\bigg\}.
\]

For each fixed $\gamma$, let
\(g(\xi\,|\gamma)\equiv\max_{\lambda\in\Lambda}\mathcal L(\xi;\gamma,\lambda)\).
Because $\Lambda$ is compact (weak topology) and $\mathcal L$ is continuous and affine, then by the generalized envelope theorem for a supremum over a compact index set (see, e.g., \citealp{bonnans2000perturbation};  \citealp{milgrom2002envelope},  the Hadamard directional derivative of $g$ at $\xi=\xi(x,y_0)$ exists and
\[
g'_H(\xi(x,y_0);\delta\, |\gamma)
=\max_{\lambda\in\Lambda^*(\gamma)}\ \bigg\{ -\gamma'_1 \delta_0\;+\;\int_{\underline y}^{\overline y}\!\big[\delta_F(y)-\gamma'_1 \delta_1(y)\big]\,d\lambda(y)\bigg\}.
\] along the direction $ \xi(x,y_0)\mapsto \xi(x,y_0)+\varepsilon\delta_\varepsilon$ as $\varepsilon\downarrow 0$.

Moreover, note that $\phi(\xi(x,y_0))=\min_{\gamma\in\mathcal B_\tau} g(\xi(x,y_0)\,|\gamma)$ with $\mathcal B_\tau$ compact and $g$ continuous in $(\xi,\gamma)$. For the \(\limsup\), take any $\gamma^*\in\Gamma^*$:
\[
\phi(\xi_\varepsilon(x,y_0))-\phi(\xi(x,y_0))
=\min_{\gamma\in\mathcal B_\tau} g(\xi_\varepsilon(x,y_0)\,|\gamma)-g(\xi(x,y_0)\, |\gamma^*)
\ \le\ g(\xi_\varepsilon(x,y_0)\, |\gamma^*)-g(\xi(x,y_0)\, |\gamma^*).
\]
It follows that 
\[
\limsup_{\varepsilon\downarrow0}\frac{\phi(\xi_\varepsilon(x,y_0))-\phi(\xi(x,y_0))}{\varepsilon}
\ \le\ g'_H(\xi(x,y_0);\delta\mid\gamma^*).
\]
Because  the above inequality holds for every $\gamma^*$ in $\Gamma^*$, taking $\min_{\gamma^*\in\Gamma^*}$ gives us
\[
\limsup_{\varepsilon\downarrow0}\frac{\phi(\xi_\varepsilon)-\phi(\xi)}{\varepsilon}
\ \le\ \min_{\gamma\in\Gamma^*}\max_{\lambda\in\Lambda^*(\gamma)}\ \bigg\{ -\gamma'_1 \delta_0\;+\;\int_{\underline y}^{\overline y}\!\big[\delta_F(y)-\gamma'_1 \delta_1(y)\big]\,d\lambda(y)\bigg\}.
\]

On the other hand, let $\gamma^*_\varepsilon\in{\arg\min}_{\gamma\in\mathcal B_\tau} g(\xi_\varepsilon(x,y_0)\, |\gamma)$.  Since $\mathcal L$ is affine in $\xi(x,y_0)$ and $\lambda$ is a probability measure,
\[
\sup_{\gamma\in\mathcal B_\tau}\Big|g(\xi_\varepsilon(x,y_0)\, |\gamma)-g(\xi(x,y_0)\, |\gamma)\Big|\leq \varepsilon\left(\|\gamma_1\|\|\delta_{0\varepsilon}\|+\|\gamma_1\|\|\delta_{1\varepsilon}(\cdot)\|_\infty+\|\delta_{F\varepsilon}(\cdot)\|_\infty\right).
\]Because $\|\gamma\|^2\leq \tau$ for each $\gamma\in\mathcal B_\tau$, 
\[
\sup_{\gamma\in\mathcal B_\tau}\Big|g(\xi_\varepsilon(x,y_0)\, |\gamma)-g(\xi(x,y_0)\, |\gamma)\Big|\rightarrow 0, \quad \text{as } \varepsilon\rightarrow 0.
\]
Consider the sequence of $\{\gamma^*_\epsilon\}_{\varepsilon\downarrow 0}$. By compactness of $\mathcal B_\tau$, there exists a sequence $\varepsilon_j\downarrow0$ such that
$\gamma_{\varepsilon_j}^*\to\bar\gamma$; by the argmin theorem (uniform convergence on a compact parameter set), $\bar\gamma\in\Gamma^*$.  Then we have 
\[
\phi(\xi_\varepsilon(x,y_0))-\phi(\xi(x,y_0))\ \ge\ g(\xi_\varepsilon(x,y_0)\, |\gamma^*_\varepsilon)-g(\xi(x,y_0)\, |\gamma^*_\varepsilon).
\]
It follows that 
\[
\liminf_{\varepsilon\downarrow0}\frac{\phi(\xi_\varepsilon(x,y_0))-\phi(\xi(x,y_0))}{\varepsilon}
\ \ge\ g'_H(\xi(x,y_0);\delta\, |\bar\gamma)
\ \ge\ \min_{\gamma\in\Gamma^*} \,  g'_H(\xi(x,y_0);\delta\, |\gamma).
\]
Combining the upper and lower bounds gives us
\[
\phi'_H(\xi(x,y_0);\delta)= \min_{\gamma\in\Gamma^*}\max_{\lambda\in\Lambda^*(\gamma)}\ \bigg\{ -\gamma'_1 \delta_0\;+\;\int_{\underline y}^{\overline y}\!\big[\delta_F(y)-\gamma'_1 \delta_1(y)\big]\,d\lambda(y)\bigg\}.\quad \qed
\]

\section{Local Stability of the Inner SILP's Optimal Solution}

By Assumption~\ref{as:R},  the inner SILP problem $\widetilde {LP}(\theta^*_2,\tau)$ should have a unique optimal solution as $ \theta^\dag_1(\theta^*_2)=\theta^*_1$. At this optimum, exact $K$ linearly independent constraints are binding and the corresponding Lagrange multipliers  are strictly positive.  Following the standard sensitivity analysis of the SILP literature, we obtain local stability of  the optimal solution. In the next lemma, we establish such a result. For $\theta_2\in\tilde B_\tau$,   we denote \( \{\lambda^\dag_k(\theta_2): k = 1, \ldots, K^\dag(\theta_2)\} \) and \( \{y^\dag_{1k}(\theta_2): k = 1, \ldots, \tilde K^\dag(\theta_2)\} \) as the Lagrange multipliers and the associated indices  for the active constraints at the optimum, respectively, where \( K^\dag(\theta_2) \in \mathbb{N} \) denotes the number of active constraints.

\begin{lemma}
\label{lem:4.4}
Suppose all the conditions in Lemmas~\ref{lem:dual},  and Assumption~\ref{as:R} hold. Then there exists an \(\varepsilon>0\) such that for all \(\theta_2\) within \(\|\theta_2-\theta_2^*\|\le\varepsilon\):  
\begin{itemize}
  \item[(i)] The number of active constraints $K^\dag(\theta_2)$ remain unchanged, i.e., $K^\dag(\theta_2)=K$, and for each $k\leq K$,  \( y_{1k}^\dagger(\theta_2)\) is continuously differentiable in $\theta_2$;
  \item [(ii)]Both \(\theta_1^\dagger(\theta_2)\) and \(\{\lambda_k^\dagger(\theta_2): k=1,\cdots,K\}\) are continuously differentiable in $\theta_2$;
  \item [(iii)]The optimal value function \(Q(\theta_2| x,y_0)\) is twice continuously differentiable in \(\theta_2\).
\end{itemize}
\proof
Consider the inner SILP problem $\widetilde {LP}(\theta^*_2,\tau)$. At the unique optimal solution $\theta_1^*$, define the normal cone as follows: 
\[
C(\theta_1^*, \theta_2^*) =  \left\{v\in \mathbb R^{K}: v'(\theta_1-\theta_1^*)\leq 0 \text { for all }(\theta_1,\theta_2^*)\in \mathbb S(x) \right\}.
\]Note that $C(\theta_1^*, \theta_2^*) \equiv C(\gamma^*) $ is generated as the cone of the derivatives of the $K$ active constraints, i.e., 
\[
C(\theta_1^*, \theta_2^*) = \left\{-\sum_{k=1}^K\mu_k[1, \, \Delta'_{11}(y^*_{1k}|x)]' :(\mu_1,\cdots,\mu_K)\in\mathbb R^K_+ \right\}.
\]Because the feasible region of $\widetilde {LP}(\theta^*_2,\tau)$ is closed and convex, and that the objective function is a linear function. Then a necessary and sufficient condition for  $\theta_1$ being an optimal solution to $\widetilde {LP}(\theta^*_2,\tau)$ is
\[
-\left[1,-\Delta'_{01}(y_0|x)\right]'\in C(\theta_1, \theta_2^*). 
\]
By Assumption~\ref{as:R}, the normal cone at $\theta_1^*$ is generated by $K$ linearly independent constraints that is smooth in the constraint index $y_1$. Moreover, because all the multipliers are strictly positive, then the negative of the objective direction, i.e., $-\left[1,-\Delta'_{01}(y_0|x)\right]$, belongs to the inner set of the normal cone $C(\theta_1, \theta_2^*)$. By \citet[][Theorem 6.43]{rockafellar1998variational}, the normal cone mapping $\theta_1 \mapsto C(\theta_1, \theta_2^*)$ is outer semicontinuous in general. Under Assumption~\ref{as:R}, the graph of the normal cone mapping is locally Lipschitz and directionally differentiable.

Hence, the geometry of the feasible region varies smoothly under small perturbations in \( \theta_1 \). Since small perturbations in \( \theta_2 \) or \( \xi( x, y_0) \) induce continuous and differentiable changes in the geometry of the feasible region, the optimal solution \( \theta_1^\dag \) responds smoothly to such perturbations. Moreover, consider the set of active constraints:
\[
 F_{Y| D, X}\big(y^\dag_{1k}(\theta_2) | 1, x\big) = \theta_{1}^\dag(\theta_2)\left[1,\, \Delta'_{11}\big(y^\dag_{1k}(\theta_2) | x\big)\right] + \theta_2' \Delta_{12}\big(y^\dag_{1k}(\theta_2) | x\big), \quad \text{for }k=1,\cdots,K.
\] By the Implicit Function Theorem, $y^\dag_{1k}(\theta_2)$ is continuously differentiable in $\theta_2$ in a small neighborhood of $\theta_2^*$, as long as $K^\dag(\theta_2)$ remains the same. Moreover, by the KKT condition:
\[
\Delta_{01}(y_0|x)=\sum_{k=1}^K\lambda^\dag_k(\theta_2)\times \Delta_{11}(y^\dag_{1k}(\theta_2)|x).
\]Again, by the implicit theorem, $\lambda^\dag_{1k}(\theta_2)$ is also continuously differentiable in in a small neighborhood of $\theta_2^*$.

We now show Condition (iii) in the lemma. By the general envelope theorem for  saddle-point problems in \citet[][Theorem 4]{milgrom2002envelope},   \(Q(\cdot| x,y_0)\) is continuously differentiable at a sufficiently small neighborhood of \(\theta^*_2\). The derivative is given by
\[
\frac{\partial Q(\theta_2 | x, y_0)}{\partial \theta_2}= - \Delta_{02}(y_0 | x) - \sum_{k=1}^K \lambda_k^\dag(\theta_2) \times \Delta_{12}\left(y^\dag_{1k}(\theta_2) | x \right).
\]
By the differentiability of $\lambda^\dag_k(\cdot)$, $\Delta_{12}(\cdot|x)$ and $y^\dag_{1k}(\cdot)$,   \(Q(\cdot| x,y_0)\) is  twice continuously differentiable at $\theta^*_2$.  \qed

\end{lemma}

In the next lemma, we apply  \citet[][]{milgrom2002envelope}'s generalized envelope theorem to  establish the Hadamard differentiability of the inner problem's value function w.r.t. $\xi(x,y_0)$. Specifically,  \cite{milgrom2002envelope}'s  Theorem 4 extends the general envelope theorem to constrained optimization over smooth manifolds, which has been ensured in our inner SILP problem under the regularity conditions  in  Assumption~\ref{as:R}. 
To proceed, let $\varphi : \mathbb{S}_{\xi(x,y_0)} \times \tilde{B}_\tau \to \mathbb{R}$ be the map from \( \xi(x,y_0) \in \mathbb{S}_{\xi(x,y_0)} \) and \( \theta_2 \in \tilde{B}_\tau \) to the optimal objective value \( Q(\theta_2 | x, y_0) \), i.e., 
\[
Q(\theta_2 | x, y_0) = \varphi(\xi(x,y_0), \theta_2).
\]By definition, $\hat Q(\cdot | x, y_0) =  \varphi(\hat \xi(x,y_0), \cdot)$, which is the sample criteria function of the outer problem. Moreover, let $\varphi'_{\xi( x, y_0)}( \delta;\theta_2) $ be the  Hadamard directional derivative of  $\varphi(\xi(x,y_0), \theta_2)$ w.r.t. $\xi(x,y_0)$ at $(\xi(x,y_0), \theta_2)$ in direction $\delta$, i.e.,
\[
\varphi'_{\xi( x, y_0)}( \delta;\theta_2)=\lim_{t\rightarrow 0}\frac{\varphi(\xi( x, y_0) +t \delta_t, \theta_2)-\varphi(\xi( x, y_0), \theta_2)}{t},
\]where $\delta_t\in \mathbb S_{\xi(x,y_0)}$ is an arbitrary sequence with $\delta_t\rightarrow \delta$  as $t\rightarrow 0$.

\begin{lemma}
\label{prop:4.1}
Suppose all the conditions in Lemmas~\ref{lem:dual} and Assumption~\ref{as:R} hold. Then, for all $\theta_2$ within $\|\theta_2-\theta_2^*\|\le \varepsilon$, where $\varepsilon>0$ is given by Lemma~\ref{lem:4.4},  functional $\varphi(\cdot,\theta_2)$ is fully Hadamard differentiable with respect to $\xi(x,y_0)$ in any direction $\delta \equiv (\delta_0,\delta_1,\delta_F)\in \mathbb S_{\xi(x,y_0)}$. Specifically,
\[
\varphi'_{\xi(x,y_0)}(\delta;\theta_2)
= - \big[\theta_1^\dag(\theta_2),\, \theta_2'\big]\delta_0(y_0| x)
+ \sum_{k=1}^K \lambda_k^\dag(\theta_2)\left\{ \delta_F\!\big(y_{1k}^\dag(\theta_2)\big)
- \big[\theta_1^\dag(\theta_2),\, \theta_2'\big]\delta_1\!\big(y_{1k}^\dag(\theta_2)\big)\right\}.
\]
Moreover, functional $\varphi(\cdot,\theta_2)$ is fully Hadamard differentiable with respect to $\theta_2$  within $\|\theta_2-\theta_2^*\|\le \varepsilon$:
\[
\varphi'_{\theta_2}(\delta;\theta_2)
=-\Psi_{02}(y_0| x)- \sum_{k=1}^K \lambda_k^\dag(\theta_2)
\Psi_{12}\!\big(y_{1k}^\dag(\theta_2)|x\big).
\]

\end{lemma}

\noindent
The result follows directly from \citet[Theorem~4]{milgrom2002envelope}. The proof is therefore omitted.


We now derive the first- and second-order derivatives of $Q(\cdot\,|x,y_0)$ with respect to $\theta_2$ in a neighborhood of $\theta_2^*$; specifically, for all $\theta_2$ such that $\|\theta_2-\theta_2^*\|\le \varepsilon$. Let  $\lambda^\dag(\theta_2)=[\lambda^\dag_1(\theta_2);\cdots; \lambda^\dag_K(\theta_2)]$ and $y_1^\dag(\theta_2)=[y_{11}^\dag(\theta_2);\cdots; y_{1K}^\dag(\theta_2)]$. By Lemma~\ref{prop:4.1}, we have
\[
  \frac{\partial Q(\theta_2|x,y_0)}{\partial \theta_2}=-\Psi_{02}(y_0|x)-\mathbb A_{12}(y^\dag_1(\theta_2)|x)\lambda^\dag(\theta_2).
\] By the KKT conditions, the primal-dual pair solution $(\theta^\dag(\theta_2), \lambda^\dag(\theta_2))$ to the inner SILP   satisfies 
 \[
 \Psi_{01}(y_0|x)=-\mathbb A'_{11}(y_1^\dag(\theta_2)|x)\lambda^\dag(\theta_2).
 \] It follows that 
\[
  \frac{\partial Q(\theta_2|x,y_0)}{\partial \theta_2}=-\Psi_{02}(y_0|x)+\mathbb A_{12}(y^\dag_1(\theta)|x)\mathbb A^{-1}_{11}(y_1^\dag(\theta_2)|x)\Psi_{01}(y_0|x).
\]
Moreover, we obtain the expression of Hessian matrix \(\frac{\partial^2 Q(\theta_2^*| x,y_0)}{\partial \theta_2\partial \theta_2'}\) as follows: 
\[
  \frac{\partial^2 Q(\theta^*_2|x,y_0)}{\partial \theta_2\partial \theta_2'}=\sum_{k=1}^K\frac{\partial [\mathbb A_{12}(y^*_1|x)\mathbb A^{-1}_{11}(y_1^*|x)]}{\partial y^*_{1k}}\Psi_{01}(y_0|x)\times \frac{\partial y^\dag_{1k}(\theta^*_2)}{\partial \theta'_2}.
\]

Thus, it suffices to obtain $\frac{\partial y^\dag_{1k}(\theta^*_2)}{\partial \theta'_2}$. Note that
\begin{equation}
\label{eq:B1}
F_{Y|DX}(y_{1k}^\dag(\theta_2)|1,x)=\Psi'_{11}(y_{1k}^\dag(\theta_2)|x)\theta^\dag_1(\theta_2)+\Psi'_{12}(y_{1k}^\dag(\theta_2)|x)\theta_2
\end{equation} for $k=1,\cdots,K$. Because each $y^\dag_{1k}(\theta_2)$ locally maximizes the  (differentiable) slack function $F_{Y|DX}(\cdot|1,x)-\Psi'_{11}(\cdot|x)\theta_1-\Psi'_{12}(\cdot|x)\theta_2$, thus the first order condition holds as follows:  
\begin{equation}
\label{eq:B2}
f_{Y|DX}\big(y_{1k}^\dag(\theta_2)|1,x\big)=\frac{\partial \Psi'_{11}(y_{1k}^\dag(\theta_2)|x)}{\partial y_{1k}}\theta_1^\dag(\theta_2)+\frac{\partial \Psi'_{12}(y_{1k}^\dag(\theta_2)|x)}{\partial y_{1k}}\theta_2
\end{equation}and the Hessian matrix is positive semi-definite. 
Combining eq.~\eqref{eq:B1} and \eqref{eq:B2}, we obtain
\[
0=\Psi'_{11}(y_{1k}^\dag(\theta_2)|x)\frac{\partial \theta^\dag_1(\theta_2)}{\partial \theta_2'}+\Psi'_{12}(y_{1k}^\dag(\theta_2)|x).
\]
Because the above condition holds for $k=1,\cdots,K$, then we have
\[
\left[\frac{\partial \theta^\dag_1(\theta_2)}{\partial \theta'_2}\right]'=-\mathbb A_{12}(y_{1}^\dag(\theta_2)|x){\mathbb A}^{-1}_{11}(y_{1}^\dag(\theta_2)|x).
\]

Moreover, we differentiate e.q.~\eqref{eq:B2} and obtain
\begin{multline*}
\left[\frac{\partial f_{Y|DX}\big(y_{1k}^\dag(\theta_2)|1,x\big)}{\partial y_{1k}}-\frac{\partial^2 \Psi'_{11}(y_{1k}^\dag(\theta_2)|x)}{\partial y^2_{1k}}\theta_1^\dag(\theta_2)-\frac{\partial^2 \Psi'_{12}(y_{1k}^\dag(\theta_2)|x)}{\partial y^2_{1k}}\theta_2\right]\frac{\partial y^\dag_{1k}(\theta_2)}{\partial \theta'_2}\\
=\frac{\partial \Psi'_{11}(y_{1k}^\dag(\theta_2)|x)}{\partial y_{1k}}\frac{\partial \theta_1^\dag(\theta_2)}{\partial \theta_2'}+\frac{\partial \Psi'_{12}(y_{1k}^\dag(\theta_2)|x)}{\partial y_{1k}}.
\end{multline*}
Denote $\Xi_k^*(x)=\frac{\partial f_{Y|DX}(y_{11}^*|1,x)}{\partial y_{11}}-\frac{\partial^2 \Psi'_{11}(y_{11}^*|x)}{\partial y^2_{11}}\theta_1^*-\frac{\partial^2 \Psi'_{12}(y_{11}^*|x)}{\partial y^2_{11}}\theta^*_2$. Note that $\Xi_k^*(x)>0$ as the second order condition for the minimization of the slack function.  Thus, we obtain
\[
\frac{\partial y^\dag_{1k}(\theta^*_2)}{\partial \theta_2}=\frac{1}{\Xi^*_k(x)} \times \mathbb V_k^*(x)
\]
where 
\begin{align*}
&\Xi_k^*(x)=\frac{\partial f_{Y|DX}(y_{11}^*|1,x)}{\partial y_{11}}-\frac{\partial^2 \Psi'_{11}(y_{11}^*|x)}{\partial y^2_{11}}\theta_1^*-\frac{\partial^2 \Psi'_{12}(y_{11}^*|x)}{\partial y^2_{11}}\theta^*_2;\\
&\mathbb V_k^*(x)=-\mathbb A_{12}(y_{1}^*|x){\mathbb A}^{-1}_{11}(y_{1}^*|x) \frac{\partial \Psi_{11}(y_{1k}^*|x)}{\partial y_{1k}}+\frac{\partial \Psi_{12}(y_{1k}^*|x)}{\partial y_{1k}}.
\end{align*}

\end{appendix}

\bibliographystyle{ecta}
\bibliography{multi_IVs}

\begin{thebibliography}{60}
\newcommand{\enquote}[1]{``#1''}
\expandafter\ifx\csname natexlab\endcsname\relax\def\natexlab#1{#1}\fi

\bibitem[\protect\citeauthoryear{Abadie, Angrist, and Imbens}{Abadie
  et~al.}{2002}]{abadie2002instrumental}
\textsc{Abadie, A., J.~Angrist, and G.~Imbens} (2002): \enquote{Instrumental
  variables estimates of the effect of subsidized training on the quantiles of
  trainee earnings,} \emph{Econometrica}, 70, 91--117.

\bibitem[\protect\citeauthoryear{Andrews}{Andrews}{1999}]{andrews1999estimation}
\textsc{Andrews, D.~W.} (1999): \enquote{Estimation when a parameter is on a
  boundary,} \emph{Econometrica}, 67, 1341--1383.

\bibitem[\protect\citeauthoryear{Billingsley}{Billingsley}{1999}]{billingsley1999convergence}
\textsc{Billingsley, P.} (1999): \emph{Convergence of Probability Measures},
  New York: John Wiley \& Sons, 2 ed.

\bibitem[\protect\citeauthoryear{Blundell, Gosling, Ichimura, and
  Meghir}{Blundell et~al.}{2007}]{blundell2007changes}
\textsc{Blundell, R., A.~Gosling, H.~Ichimura, and C.~Meghir} (2007):
  \enquote{Changes in the distribution of male and female wages accounting for
  employment composition using bounds,} \emph{Econometrica}, 75, 323--363.

\bibitem[\protect\citeauthoryear{Bonnans and Shapiro}{Bonnans and
  Shapiro}{2000}]{bonnans2000perturbation}
\textsc{Bonnans, J.~F. and A.~Shapiro} (2000): \emph{Perturbation Analysis of
  Optimization Problems}, Springer Series in Operations Research, New York:
  Springer.

\bibitem[\protect\citeauthoryear{Buchholz, Shum, and Xu}{Buchholz
  et~al.}{2021}]{buchholz2021semiparametric}
\textsc{Buchholz, N., M.~Shum, and H.~Xu} (2021): \enquote{Semiparametric
  estimation of dynamic discrete choice models,} \emph{Journal of
  Econometrics}, 223, 312--327.

\bibitem[\protect\citeauthoryear{Chernozhukov, Fern{\'a}ndez-Val, and
  Galichon}{Chernozhukov et~al.}{2010}]{chernozhukov2010quantile}
\textsc{Chernozhukov, V., I.~Fern{\'a}ndez-Val, and A.~Galichon} (2010):
  \enquote{Quantile and probability curves without crossing,}
  \emph{Econometrica}, 78, 1093--1125.

\bibitem[\protect\citeauthoryear{Chernozhukov and Hansen}{Chernozhukov and
  Hansen}{2005}]{chernozhukov2005iv}
\textsc{Chernozhukov, V. and C.~Hansen} (2005): \enquote{An {IV} model of
  quantile treatment effects,} \emph{Econometrica}, 73, 245--261.

\bibitem[\protect\citeauthoryear{Chernozhukov and Hansen}{Chernozhukov and
  Hansen}{2013}]{ChernozhukovHansen2013ARE}
---\hspace{-.1pt}---\hspace{-.1pt}--- (2013): \enquote{Quantile Models with
  Endogeneity,} \emph{Annual Review of Economics}, 5, 57--81.

\bibitem[\protect\citeauthoryear{Chernozhukov, Hong, and Tamer}{Chernozhukov
  et~al.}{2007}]{chernozhukov2007estimation}
\textsc{Chernozhukov, V., H.~Hong, and E.~Tamer} (2007): \enquote{Estimation
  and confidence regions for parameter sets in econometric models 1,}
  \emph{Econometrica}, 75, 1243--1284.

\bibitem[\protect\citeauthoryear{Chesher}{Chesher}{2003}]{chesher2003identification}
\textsc{Chesher, A.} (2003): \enquote{Identification in nonseparable models,}
  \emph{Econometrica}, 71, 1405--1441.

\bibitem[\protect\citeauthoryear{Chesher}{Chesher}{2005}]{Che05}
---\hspace{-.1pt}---\hspace{-.1pt}--- (2005): \enquote{Nonparametric
  identification under discrete variation,} \emph{Econometrica}, 73,
  1525--1550.

\bibitem[\protect\citeauthoryear{Christensen and Connault}{Christensen and
  Connault}{2023}]{ChristensenConnault2023}
\textsc{Christensen, T.~M. and B.~Connault} (2023): \enquote{Counterfactual
  Sensitivity and Robustness,} \emph{Econometrica}, 91, 1695--1736.

\bibitem[\protect\citeauthoryear{de~Castro}{de~Castro}{2007}]{Castro2007}
\textsc{de~Castro, L.~I.} (2007): \enquote{Affiliation, Equilibrium Existence
  and the Revenue Ranking of Auctions,} Working papers.

\bibitem[\protect\citeauthoryear{D'Haultf{\oe}uille and
  F{\'e}vrier}{D'Haultf{\oe}uille and F{\'e}vrier}{2015}]{d2015identification}
\textsc{D'Haultf{\oe}uille, X. and P.~F{\'e}vrier} (2015):
  \enquote{Identification of nonseparable triangular models with discrete
  instruments,} \emph{Econometrica}, 83, 1199--1210.

\bibitem[\protect\citeauthoryear{Dong and Shen}{Dong and
  Shen}{2018}]{dong2018testing}
\textsc{Dong, Y. and S.~Shen} (2018): \enquote{Testing for rank invariance or
  similarity in program evaluation,} \emph{Review of Economics and Statistics},
  100, 78--85.

\bibitem[\protect\citeauthoryear{Ekeland and Temam}{Ekeland and
  Temam}{1999}]{ekeland1999convex}
\textsc{Ekeland, I. and R.~Temam} (1999): \emph{Convex Analysis and Variational
  Problems}, SIAM, reprint of the 1976 english edition ed.

\bibitem[\protect\citeauthoryear{Fang and Santos}{Fang and
  Santos}{2019}]{fang2019inference}
\textsc{Fang, Z. and A.~Santos} (2019): \enquote{Inference on directionally
  differentiable functions,} \emph{The Review of Economic Studies}, 86,
  377--412.

\bibitem[\protect\citeauthoryear{Fang, Santos, Shaikh, and Torgovitsky}{Fang
  et~al.}{2023}]{fang2023inference}
\textsc{Fang, Z., A.~Santos, A.~M. Shaikh, and A.~Torgovitsky} (2023):
  \enquote{Inference for Large-Scale Linear Systems With Known Coefficients,}
  \emph{Econometrica}, 91, 299--327.

\bibitem[\protect\citeauthoryear{Fiacco}{Fiacco}{1983}]{fiacco1983introduction}
\textsc{Fiacco, A.~V.} (1983): \enquote{Introduction to sensitivity and
  stability analysis in non linear programming,} .

\bibitem[\protect\citeauthoryear{Foresi and Peracchi}{Foresi and
  Peracchi}{1995}]{foresi1995conditional}
\textsc{Foresi, S. and F.~Peracchi} (1995): \enquote{The conditional
  distribution of excess returns: An empirical analysis,} \emph{Journal of the
  American Statistical Association}, 90, 451--466.

\bibitem[\protect\citeauthoryear{Frandsen and Lefgren}{Frandsen and
  Lefgren}{2018}]{frandsen2018testing}
\textsc{Frandsen, B.~R. and L.~J. Lefgren} (2018): \enquote{Testing rank
  similarity,} \emph{Review of Economics and Statistics}, 100, 86--91.

\bibitem[\protect\citeauthoryear{Goberna and L{\'o}pez}{Goberna and
  L{\'o}pez}{1998{\natexlab{a}}}]{goberna1998comprehensive}
\textsc{Goberna, M.~A. and M.~A. L{\'o}pez} (1998{\natexlab{a}}): \enquote{A
  comprehensive survey of linear semi-infinite optimization theory,}
  \emph{Semi-infinite programming}, 3--27.

\bibitem[\protect\citeauthoryear{Goberna and L{\'o}pez}{Goberna and
  L{\'o}pez}{1998{\natexlab{b}}}]{goberna1998linear}
---\hspace{-.1pt}---\hspace{-.1pt}--- (1998{\natexlab{b}}): \emph{Linear
  Semi-Infinite Optimization}, Wiley-Interscience.

\bibitem[\protect\citeauthoryear{Goff and Mbakop}{Goff and
  Mbakop}{2025}]{GoffMbakop2025}
\textsc{Goff, L. and E.~D. Mbakop} (2025): \enquote{Inference on the Value of a
  Linear Program,} Working paper.

\bibitem[\protect\citeauthoryear{Han}{Han}{2021}]{han2021identification}
\textsc{Han, S.} (2021): \enquote{Identification in nonparametric models for
  dynamic treatment effects,} \emph{Journal of Econometrics}, 225, 132--147.

\bibitem[\protect\citeauthoryear{Han and Yang}{Han and
  Yang}{2023}]{han2020sharp}
\textsc{Han, S. and S.~Yang} (2023): \enquote{A Computational Approach to
  Identification of Treatment Effects for Policy Evaluation,} \emph{arXiv
  preprint arXiv:2009.13861}.

\bibitem[\protect\citeauthoryear{H{\"a}rdle, Werwatz, M{\"u}ller, and
  Sperlich}{H{\"a}rdle et~al.}{2004}]{hardle2004nonparametric}
\textsc{H{\"a}rdle, W., A.~Werwatz, M.~M{\"u}ller, and S.~Sperlich} (2004):
  \enquote{Nonparametric regression,} in \emph{Nonparametric and Semiparametric
  Models}, Springer, 85--141.

\bibitem[\protect\citeauthoryear{Heckman and Robb}{Heckman and
  Robb}{1986}]{heckman1986alternative}
\textsc{Heckman, J.~J. and R.~Robb} (1986): \enquote{Alternative methods for
  solving the problem of selection bias in evaluating the impact of treatments
  on outcomes,} in \emph{Drawing inferences from self-selected samples},
  Routledge, 63--107.

\bibitem[\protect\citeauthoryear{Heckman, Smith, and Clements}{Heckman
  et~al.}{1997}]{heckman1997making}
\textsc{Heckman, J.~J., J.~Smith, and N.~Clements} (1997): \enquote{Making the
  most out of programme evaluations and social experiments: Accounting for
  heterogeneity in programme impacts,} \emph{The Review of Economic Studies},
  64, 487--535.

\bibitem[\protect\citeauthoryear{Hettich and Kortanek}{Hettich and
  Kortanek}{1993}]{hettich1993}
\textsc{Hettich, R. and K.~O. Kortanek} (1993): \enquote{Semi-infinite
  programming: Theory, methods, and applications,} \emph{SIAM Review}, 35,
  380--429.

\bibitem[\protect\citeauthoryear{Hong and Li}{Hong and
  Li}{2018}]{hong2018numerical}
\textsc{Hong, H. and J.~Li} (2018): \enquote{The Numerical Delta Method,}
  \emph{Journal of Econometrics}, 206, 379--394.

\bibitem[\protect\citeauthoryear{Hornik, Stinchcombe, and White}{Hornik
  et~al.}{1989}]{hornik1989multilayer}
\textsc{Hornik, K., M.~Stinchcombe, and H.~White} (1989): \enquote{Multilayer
  feedforward networks are universal approximators,} \emph{Neural networks}, 2,
  359--366.

\bibitem[\protect\citeauthoryear{Hsieh, Shi, and Shum}{Hsieh
  et~al.}{2022}]{hsieh2022inference}
\textsc{Hsieh, Y.-W., X.~Shi, and M.~Shum} (2022): \enquote{Inference on
  Estimators Defined by Mathematical Programming,} \emph{Econometrica}, 90,
  661--691.

\bibitem[\protect\citeauthoryear{Imbens and Angrist}{Imbens and
  Angrist}{1994}]{imbens1994identification}
\textsc{Imbens, G.~W. and J.~D. Angrist} (1994): \enquote{Identification and
  Estimation of Local Average Treatment Effects,} \emph{Econometrica}, 62,
  467--475.

\bibitem[\protect\citeauthoryear{Imbens and Rubin}{Imbens and
  Rubin}{1997}]{imbens1997estimating}
\textsc{Imbens, G.~W. and D.~B. Rubin} (1997): \enquote{Estimating outcome
  distributions for compliers in instrumental variables models,} \emph{The
  Review of Economic Studies}, 64, 555--574.

\bibitem[\protect\citeauthoryear{Jun, Pinkse, and Xu}{Jun
  et~al.}{2011}]{jun2011tighter}
\textsc{Jun, S.~J., J.~Pinkse, and H.~Xu} (2011): \enquote{Tighter bounds in
  triangular systems,} \emph{Journal of Econometrics}, 161, 122--128.

\bibitem[\protect\citeauthoryear{Kim and Park}{Kim and
  Park}{2022}]{kim2022testing}
\textsc{Kim, J.~H. and B.~G. Park} (2022): \enquote{Testing rank similarity in
  the local average treatment effects model,} \emph{Econometric Reviews},
  1--22.

\bibitem[\protect\citeauthoryear{Maasoumi and Wang}{Maasoumi and
  Wang}{2019}]{maasoumi2019gender}
\textsc{Maasoumi, E. and L.~Wang} (2019): \enquote{The gender gap between
  earnings distributions,} \emph{Journal of Political Economy}, 127,
  2438--2504.

\bibitem[\protect\citeauthoryear{Manski}{Manski}{1990}]{manski1990nonparametric}
\textsc{Manski, C.~F.} (1990): \enquote{Nonparametric bounds on treatment
  effects,} \emph{The American Economic Review}, 80, 319--323.

\bibitem[\protect\citeauthoryear{Manski}{Manski}{1994}]{manski1994selection}
---\hspace{-.1pt}---\hspace{-.1pt}--- (1994): \enquote{The selection problem,}
  in \emph{Advances in Econometrics, Sixth World Congress, ed. by C. Sims},
  vol.~1, 143--70.

\bibitem[\protect\citeauthoryear{Manski}{Manski}{1997}]{manski1997monotone}
---\hspace{-.1pt}---\hspace{-.1pt}--- (1997): \enquote{Monotone treatment
  response,} \emph{Econometrica: Journal of the Econometric Society},
  1311--1334.

\bibitem[\protect\citeauthoryear{Manski and Pepper}{Manski and
  Pepper}{2000}]{MP00}
\textsc{Manski, C.~F. and J.~V. Pepper} (2000): \enquote{Monotone instrumental
  variables: With an application to the returns to schooling,}
  \emph{Econometrica}, 68, 997--1010.

\bibitem[\protect\citeauthoryear{Milgrom and Segal}{Milgrom and
  Segal}{2002}]{milgrom2002envelope}
\textsc{Milgrom, P. and I.~Segal} (2002): \enquote{Envelope theorems for
  arbitrary choice sets,} \emph{Econometrica}, 70, 583--601.

\bibitem[\protect\citeauthoryear{Mogstad, Santos, and Torgovitsky}{Mogstad
  et~al.}{2018}]{mogstad2018using}
\textsc{Mogstad, M., A.~Santos, and A.~Torgovitsky} (2018): \enquote{Using
  instrumental variables for inference about policy relevant treatment
  parameters,} \emph{Econometrica}, 86, 1589--1619.

\bibitem[\protect\citeauthoryear{Mogstad, Torgovitsky, and Walters}{Mogstad
  et~al.}{2021}]{mogstad2021causal}
\textsc{Mogstad, M., A.~Torgovitsky, and C.~R. Walters} (2021): \enquote{The
  causal interpretation of two-stage least squares with multiple instrumental
  variables,} \emph{American Economic Review}, 111, 3663--98.

\bibitem[\protect\citeauthoryear{Newey and McFadden}{Newey and
  McFadden}{1994}]{newey1994large}
\textsc{Newey, W.~K. and D.~McFadden} (1994): \enquote{Large Sample Estimation
  and Hypothesis Testing,} in \emph{Handbook of Econometrics}, ed. by R.~F.
  Engle and D.~McFadden, Elsevier, vol.~4, 2111--2245.

\bibitem[\protect\citeauthoryear{Pagan and Ullah}{Pagan and
  Ullah}{1999}]{paganullah1999}
\textsc{Pagan, A. and A.~Ullah} (1999): \emph{Nonparametric Econometrics},
  Cambridge University Press.

\bibitem[\protect\citeauthoryear{Pomatto, Strack, and Tamuz}{Pomatto
  et~al.}{2020}]{pomatto2020stochastic}
\textsc{Pomatto, L., P.~Strack, and O.~Tamuz} (2020): \enquote{Stochastic
  dominance under independent noise,} \emph{Journal of Political Economy}, 128,
  1877--1900.

\bibitem[\protect\citeauthoryear{Rockafellar}{Rockafellar}{1970}]{rockafellar1970convex}
\textsc{Rockafellar, R.~T.} (1970): \emph{Convex Analysis}, Princeton, NJ:
  Princeton University Press.

\bibitem[\protect\citeauthoryear{Rockafellar and Wets}{Rockafellar and
  Wets}{1998}]{rockafellar1998variational}
\textsc{Rockafellar, R.~T. and R.~J. Wets} (1998): \emph{Variational analysis},
  Springer.

\bibitem[\protect\citeauthoryear{Rubin}{Rubin}{1974}]{rubin1974estimating}
\textsc{Rubin, D.~B.} (1974): \enquote{Estimating causal effects of treatments
  in randomized and nonrandomized studies.} \emph{Journal of Educational
  Psychology}, 66, 688.

\bibitem[\protect\citeauthoryear{Shaikh and Vytlacil}{Shaikh and
  Vytlacil}{2011}]{SV11}
\textsc{Shaikh, A.~M. and E.~J. Vytlacil} (2011): \enquote{Partial
  identification in triangular systems of equations with binary dependent
  variables,} \emph{Econometrica}, 79, 949--955.

\bibitem[\protect\citeauthoryear{Shapiro}{Shapiro}{1991}]{shapiro1991asymptotic}
\textsc{Shapiro, A.} (1991): \enquote{Asymptotic analysis of stochastic
  programs,} \emph{Annals of Operations Research}, 30, 169--186.

\bibitem[\protect\citeauthoryear{Sion}{Sion}{1958}]{sion1958general}
\textsc{Sion, M.} (1958): \enquote{On General Minimax Theorems,} \emph{Pacific
  Journal of Mathematics}, 8, 171--176.

\bibitem[\protect\citeauthoryear{Torgovitsky}{Torgovitsky}{2015}]{torgovitsky2015identification}
\textsc{Torgovitsky, A.} (2015): \enquote{Identification of nonseparable models
  using instruments with small support,} \emph{Econometrica}, 83, 1185--1197.

\bibitem[\protect\citeauthoryear{van~der Vaart}{van~der
  Vaart}{1998}]{vandervaart1998asymptotic}
\textsc{van~der Vaart, A.~W.} (1998): \emph{Asymptotic Statistics}, Cambridge
  Series in Statistical and Probabilistic Mathematics, Cambridge University
  Press.

\bibitem[\protect\citeauthoryear{van~der Vaart and Wellner}{van~der Vaart and
  Wellner}{1996}]{vandervaart1996weak}
\textsc{van~der Vaart, A.~W. and J.~A. Wellner} (1996): \emph{Weak Convergence
  and Empirical Processes: With Applications to Statistics}, Springer Series in
  Statistics, New York: Springer.

\bibitem[\protect\citeauthoryear{Vuong and Xu}{Vuong and
  Xu}{2017}]{vuong2017counterfactual}
\textsc{Vuong, Q. and H.~Xu} (2017): \enquote{Counterfactual mapping and
  individual treatment effects in nonseparable models with binary endogeneity,}
  \emph{Quantitative Economics}, 8, 589--610.

\bibitem[\protect\citeauthoryear{Vytlacil and Yildiz}{Vytlacil and
  Yildiz}{2007}]{VY07}
\textsc{Vytlacil, E. and N.~Yildiz} (2007): \enquote{Dummy endogenous variables
  in weakly separable models,} \emph{Econometrica}, 75, 757--779.

\end{thebibliography}

\end{document}